\pdfoutput=1
\documentclass[12pt]{article}
\usepackage{amsmath,amssymb,array,calc,rotating,epsfig,psfrag,amscd, cite}
\usepackage{booktabs}
\usepackage{hyperref}
\usepackage{epsfig}
\usepackage{amsmath}
\usepackage{subcaption,tikz}
\usetikzlibrary{shapes}
\usetikzlibrary{plotmarks}

\makeatletter
\renewcommand\section{\@startsection {section}{1}{\z@}%
                                   {-3.5ex \@plus -1ex \@minus -.2ex}
                                   {2.3ex \@plus.2ex}%
                                   {\normalfont\large\bfseries}}
\renewcommand\subsection{\@startsection{subsection}{2}{\z@}%
                                     {-3.25ex\@plus -1ex \@minus -.2ex}%
                                     {1.5ex \@plus .2ex}%
                                     {\normalfont\bfseries}}
\makeatother

\newcommand{\be}{\begin{equation}}
\newcommand{\ee}{\end{equation}}
\newcommand{\bea}{\begin{eqnarray}}
\newcommand{\eea}{\end{eqnarray}}

\newcommand{\al}{\alpha}
\renewcommand{\d}{\delta}

\renewcommand{\k}{\kappa}
\newcommand{\La}{\Lambda}

\newcommand{\m}{\mu}
\newcommand{\n}{\nu}

\newcommand{\s}{\sigma}

\newcommand{\hlf}{\frac{1}{2}}

\newcommand{\non}{\nonumber}

\newcommand{\p}{\partial}

\newcommand{\rr}{\rightarrow}

\newcommand{\Z}{\mathbb{Z}}

\newcommand{\Spin}{\operatorname{Spin}}
\newcommand{\SU}{\operatorname{SU}}

\newcommand{\lp}{\left(}
\newcommand{\rp}{\right)}
\newcommand{\ls}{\left[}
\newcommand{\rs}{\right]}

\newcommand{\SW}{\mathcal{SW}}

\newcommand{\wtL}{{\widetilde{\Lambda}}}

\newcommand{\wtU}{{\widetilde{U}}}

\begin{document}
\begin{titlepage}

\begin{center}

\hfill January 9, 2024

\vskip 2 cm
{\Large \bf Characters and relations among $\SW(3/2,2)$ algebras}\\
\vskip 1.25 cm {Daniel Robbins and Chris Simmons}\\

{\vskip 0.5cm \it Depatment of Physics, University at Albany, \\ Albany, NY 12222, USA \\}

\end{center}
\vskip 2 cm

\begin{abstract}
\baselineskip=18pt
The $\operatorname{\SW}(3/2,2)$ current algebras come in two discrete series indexed by central charge, with the chiral algebra of a supersymmetric sigma model on a $\operatorname{Spin}(7)$ manifold as a special case.  The unitary representations of these algebras were classified by Gepner and Noyvert, and we use their results to perform an analysis of null descendants and compute the characters for every representation.  We obtain threshold relations between the characters of discrete representations and those with continuous conformal weights.  Modular transformations are discussed, and we show that the continuous characters can be written as bilinear combinations of characters for consecutive minimal models.

\end{abstract}

\end{titlepage}

\newpage

\tableofcontents

\newpage

\pagestyle{plain}
\baselineskip=19pt
\section{Introduction}

String theory has illuminated many fascinating connections between physics and mathematics.  One class of examples which motivates this work comes from compactifications of strings on special holonomy manifolds.  The most familiar examples are Calabi-Yau manifolds with holonomy $\SU(n)$.  The worldsheet theory of a string\footnote{We have in mind a type II superstring whose basic definition includes $(1,1)$ supersymmetry on the worldsheet, but heterotic strings also furnish interesting examples.} moving in such a background includes a sigma model with the Calabi-Yau target space.  The $\SU(n)$ holonomy ensures that there are additional conserved currents in the worldsheet theory.  The K\"ahler form leads to an extra spin one current that enhances the worldsheet supersymmetry from $(1,1)$ to $(2,2)$, while the holomorphic $n$-form adds an additional spin $n/2$ current which relates to spectral flow~\cite{Odake:1989dm}.

In the Calabi-Yau case the connections between the geometry, the worldsheet theory, and the effective theory have been well explored.  The multiplicities of discrete representations in the superconformal sigma model are determined by the Hodge numbers of the Calabi-Yau, and give rise to the BPS states in the spectrum of the effective theory.  The spectrum of continuous representations is related to the spectra of Laplace-type operators on the manifold, and to massive states in the effective theory, while OPE coefficients in the sigma model relate to couplings in the effective theory and, in some cases, to intersection numbers on the Calabi-Yau.

These connections are much less well-studied for the cases of exceptional holonomy manifolds, where the holonomy group is either $G_2$ in seven dimensions, or $\Spin(7)$ in eight dimensions.  In particular, Shatashvili and Vafa~\cite{Shatashvili:1994zw} observed that superconformal sigma models with exceptional holonomy target spaces would have their chiral algebras enhanced by additional higher spin currents.  In the $G_2$ case one obtains an $\SW(3/2,3/2,2)$ theory with central charge $c=21/2$, and in the $\Spin(7)$ case one obtains an $\SW(3/2,2)$ theory with central charge $c=12$.  In both cases these algebras are particular members of infinite discrete families with different central charges~\cite{Figueroa-OFarrill:1990tqt,Figueroa-OFarrill:1990mzn,Figueroa-OFarrill:1996tnk,Gepner:2001px}.  

For the $c=21/2$ and $c=12$ cases, one can again find correspondences between discrete representations of the CFT, topological invariants of the manifold, and BPS states in the effective theory~\cite{Shatashvili:1994zw}, and one expects that in principle the other data of the CFT will correspond to geometric information, and will map to effective theory data, in ways that are analogous to the Calabi-Yau case, though the details have not been extensively explored.  

A preliminary step towards studying these issues is to elucidate the representation theory of these chiral algebras.  A very large step in that direction was taken by Gepner and Noyvert in~\cite{Gepner:2001px}, where they classified the unitary representations for all members of the discrete $\SW(3/2,2)$ family of theories, with a similar classification undertaken by Noyvert~\cite{Noyvert:2002mc} for the case of the $\SW(3/2,3/2,2)$ theories.  These works found the representations but did not compute their characters.  For the $c=12$ theory that is relevant for $\Spin(7)$ compactifications, expressions for the continuous characters were found in~\cite{Eguchi:2003yy} and expressions for all the $c=12$ characters were conjectured (with much evidence presented) in~\cite{Benjamin:2014kna}.  Other somewhat recent work on these algebras include~\cite{Ferrari:2017kbp,Fiset:2019ecu,Fiset:2020lmg,Fiset:2021ruv,GaldeanoSolans:2022mki}.

What about the other members of the discrete families?  Very few examples of constructed theories with these algebras are known (see~\cite{Naka:2002xs} for a suggestion on coset constructions for the lower series models), but it is tempting to speculate that one could engineer instances of them and potentially use them as building blocks to give new, non-perturbative constructions of $G_2$ or $\Spin(7)$ string compactifications, in analogy with the Gepner models~\cite{Gepner:1987qi} which build Calabi-Yau compactifications as orbifolds of products of $\mathcal{N}=2$ minimal models\footnote{For related work constructing $G_2$ holonomy theories as orbifolds of $\mathcal{N}=2$ Gepner models, see e.g.~\cite{Eguchi:2001ip,Roiban:2001cp}.}.  However, before we can realize such hopes, we need more examples of these theories and a better understanding of their structure, such as their spectra and OPE coefficients.  Absent other explicit constructions, one approach for bounding, or perhaps even isolating, these theories, is via bootstrap methods.  The modular bootstrap~\cite{Hellerman:2009bu}, specifically an approach along the lines of~\cite{Keller:2012mr} or~\cite{Afkhami-Jeddi:2017idc}, might be able to give us useful information about such CFTs.

The current paper is a first step towards implementing such a program for the other members of the $\SW(3/2,2)$ family of theories.  We will work to construct the characters for the unitary representations that were found by~\cite{Gepner:2001px}.  By using an embedding diagram formalism, we will obtain explicit expressions for these characters in both the continuous and discrete representations.  As in~\cite{Benjamin:2014kna}, we do not rigorously prove that we are correctly accounting for all relations among singular states, so our expressions are still, in some sense, conjectural.  However, we will show that they pass several consistency checks.  In the case of the continuous representations we have an alternative approach to obtaining the characters by appealing to modular invariance and the role played by consecutive pairs of unitary minimal models.  Namely we find that besides the manifest appearance of a unitary bosonic minimal model in the structure of each $\SW(3/2,2)$ algebra (the minimal model Virasoro algebra appears as a subalgebra of the theory's chiral algebra), the massive characters also carry the structure of a second bosonic minimal model with a consecutive index.  For example the $c=12$ theory has an Ising model Virasoro subalgebra ($c=\hlf$), but the massive characters can also be decomposed as representations of the tri-critical Ising model ($c=\frac{7}{10}$).

For ease of reference, we include the full expressions for the characters in Tables~\ref{tab:USChars} and ~\ref{tab:LSChars}.  Table~\ref{tab:USChars} contains the expressions for the upper series.  For the continuous representations, in either the NS or R sectors, they are most conveniently written as a bilinear combination of minimal model characters, whose expressions are given in~(\ref{eq:MMChars}).  For the discrete characters we have a general template as a sum over $k\in\Z$, where the summand is a universal piece $q^{\hlf p(p+2)k^2}$ times a representation-dependent piece $\xi_k$ which is listed for each case below.  The pre-factors $P_{NS}$ or $P_R$ are the contributions for descendants of a free algebra, and are given in~(\ref{eq:PNS}) and~(\ref{eq:PR}).  For more details of the various representations, including their weights and charges, the reader should consult section~\ref{sec:Representations}.

The paper is organized as follows.  In section~\ref{sec:Algebras} we briefly review the structure of the $\SW(3/2,2)$ algebras and Gepner and Noyvert's classification of the possible unitary realizations and in section~\ref{sec:Representations} we review their classification of representations.  Section~\ref{sec:Embedding} establishes our approach to the embedding formalism.  Section~\ref{sec:ContinuousRepCharacters} computes the characters for the continuous representations using the embedding formalism.  We then also conjecture a rewriting of the characters (which we can check numerically to high order in $q$) as bilinear products of characters from $p$'th and $(p+1)$'th minimal models, and show that the results have surprisingly nice modular properties which essentially fix them uniquely.  Section~\ref{sec:ThresholdRelations} takes a detour to discuss the threshold relations that describe how the continuous (massive) representations decompose into discrete representations.  These relations will provide powerful checks on our results.  Section~\ref{sec:DiscreteRepCharacters} then computes the discrete characters using the embedding formalism.  As one last check, section~\ref{sec:SpecialCases} compares our expressions for $c=3/2$ with the corresponding expressions for the supersymmetric free boson and related theories.

\begin{table}
\centering
{\setlength{\extrarowheight}{3mm}
\begin{tabular}{|l|l|c|}
\hline
\multicolumn{3}{|l|}{{\bf{Upper series, }}\ $c=6+\frac{18}{p}$,\ $p\ge 3$} \\[3mm]
\hline
\multicolumn{3}{|l|}{{\bf{Continuous reps:}}\quad $\chi[U^{(p)}_{a,b;x}](q)=\frac{q^{x-y_{a,b}}}{\eta(q)}\sum_{k=1}^p\chi^{(p+1)}_{k,a}(q)\chi^{(p)}_{b,k}(q),$}\\
\multicolumn{3}{|l|}{$1\le a\le p+1,\ 1\le b\le p-1,\ x>0,\ \quad y_{a,b}=\frac{\lp p+2-2a\rp^2}{8p(p+2)}+\frac{\d_{a,1}+\d_{a,p+1}}{p}.$} \\[3mm]
\hline
\multicolumn{2}{|l|}{{\bf{Discrete reps:}}} & $\xi_k$ \\[3mm]
\hline
\multicolumn{3}{|l|}{{\bf{NS}}, \quad $\chi=P_{NS}(q)q^{h-\frac{c}{24}}\sum_{k\in\Z}q^{\hlf p(p+2)k^2}\xi_k$} \\[3mm]
\hline
$A$ & & $\frac{q^{(p+1)k-\hlf}}{1+q^{pk-\hlf}}-\frac{q^{(p+1)k+\hlf}}{1+q^{pk+\frac{3}{2}}}$ \\[3mm]
\hline
$B_n$ & $1\le n\le\frac{p-2}{2}$ & $\frac{q^{((p+2)n+p+1)k}}{1+q^{pk+n-\hlf}}-\frac{q^{((p+2)n+p+1)k+1}}{1+q^{pk+n+\frac{3}{2}}}$ \\[3mm]
\hline
$C_{n,m}$ & $1\le n\le p-2,\ n+1\le m\le\frac{p+n}{2}$ & $\frac{q^{(pm-(p+1)n)k}}{1+q^{pk+m-n-\hlf}}-\frac{q^{(pm+n)k+\hlf n(2m-n)}}{1+q^{pk+m-\hlf}}$ \\[3mm]
\hline
$D_{n,m}$ & $2\le n\le p-1,\ \frac{n}{2}\le m\le n-1$ & $\frac{q^{((p+1)n-pm)k}}{1+q^{pk+n-m-\hlf}}-\frac{q^{(pm+n+p)k+\hlf (n+1)(2m-n+1)}}{1+q^{pk+m-\hlf}}$ \\[3mm]
\hline
\multicolumn{3}{|l|}{{\bf{R}}, \quad $\chi=P_R(q)q^{h-\frac{c}{24}}\sum_{k\in\Z}q^{\hlf p(p+2)k^2}\xi_k$} \\[3mm]
\hline
$E$ & & $\frac{q^{\hlf(3p+4)k}}{1+q^{pk}}-\frac{q^{\hlf(3p+4)k+1}}{1+q^{pk+2}}$ \\[3mm]
\hline
$F_n$ & $1\le n\le p-1$ & $\frac{q^{\hlf(2n+p)k}}{1+q^{pk}}-\frac{q^{((p+1)n+\frac{p}{2})k+\hlf n(n+1)}}{1+q^{pk+n}}$ \\[3mm]
\hline
$G_{n,m}$ & $1\le n\le p-3,\ n+1\le m\le\frac{p-1+n}{2}$ & $\frac{q^{(pm-(p+1)n+\frac{p}{2})k}}{1+q^{pk+m-n}}-\frac{q^{(pm+n+\frac{p}{2})k+\hlf n(2m-n+1)}}{1+q^{pk+m}}$ \\[3mm]
\hline
$H_{n,m}$ & $1\le n\le p-3,\ n+2\le m\le\frac{p+1+n}{2}$ & $\frac{q^{(pm-(p+1)n+\frac{p}{2})k}}{1+q^{pk+m-n-1}}-\frac{q^{(pm+n+\frac{p}{2})k+\hlf n(2m-n+1)}}{1+q^{pk+m-1}}$ \\[3mm]
\hline
$I_n$ & $2\le n\le\frac{p-1}{2}$ & $\frac{q^{((p+2)n+\frac{p}{2})k}}{1+q^{pk+n-1}}-\frac{q^{((p+2)n+\frac{p}{2})k+1}}{1+q^{pk+n+1}}$ \\[3mm]
\hline
\end{tabular}}
\caption{Characters for representations of the upper series.  Here $p$, $a$, $b$, $n$, and $m$ are integers, and $x$ is a positive real number.}
\label{tab:USChars}
\end{table}

\begin{table}
\centering
{\setlength{\extrarowheight}{3mm}
\begin{tabular}{|l|l|c|}
\hline
\multicolumn{2}{|l|}{{\bf{Lower series, }}\ $c=6-\frac{18}{p+1}$,\ $p\ge 3$} & $\xi_k$ \\[3mm]
\hline
\multicolumn{3}{|l|}{{\bf{Continuous reps: }}\quad $\chi[\wtU^{(p)}_{a,b;x}](q)=\frac{q^{x-y'_{a,b}}}{\eta(q)}\sum_{k=1}^{p-1}\chi^{(p-1)}_{b,k}(q)\chi^{(p)}_{k,a}(q),$}\\
\multicolumn{3}{|l|}{$1\le a\le p,\ 1\le b\le p-2,\ x>0,\ \quad y'_{a,b}=\frac{(p-1-2b)^2}{8(p-1)(p+1)}.$} \\[3mm]
\hline
\multicolumn{2}{|l|}{{\bf{Discrete reps: }}} & $\xi_k$ \\[3mm]
\hline
\multicolumn{3}{|l|}{{\bf{NS}}, \quad $\chi=P_{NS}(q)q^{h-\frac{c}{24}}\sum_{k\in\Z}q^{\hlf(p-1)(p+1)k^2}\xi_k$} \\[3mm]
\hline
$\widetilde{A}_{n,m}$ & $1\le m\le p-1,\ m\le n\le\frac{p-1+m}{2}$ & $\frac{q^{((p+1)n-pm)k}}{1+q^{(p+1)k+n-m+\hlf}}-\frac{q^{((p+1)n-m)k+\hlf m(2n-m)}}{1+q^{(p+1)k+n+\hlf}}$ \\[3mm]
\hline
$\widetilde{B}_{n,m}$ & $1\le m\le p-2,\ m+1\le n\le\frac{p+m}{2}$ & $\frac{q^{((p+1)n-pm)k}}{1+q^{(p+1)k+n-m-\hlf}}-\frac{q^{((p+1)n-m)k+\hlf m(2n-m)}}{1+q^{(p+1)k+n-\hlf}}$ \\[3mm]
\hline
\multicolumn{3}{|l|}{{\bf{R}}, \quad $\chi=P_R(q)q^{h-\frac{c}{24}}\sum_{k\in\Z}q^{\hlf(p-1)(p+1)k^2}\xi_k$} \\[3mm]
\hline
$\widetilde{C}_n$ & $1\le n\le p-1$ & $\frac{q^{(n+\frac{p+1}{2})k}}{1+q^{(p+1)k}}-\frac{q^{(pn+\frac{p+1}{2})k+\hlf n(n+1)}}{1+q^{(p+1)k+n}}$ \\[3mm]
\hline
$\widetilde{D}_{n,m}$ & $1\le m\le p-3,\ m+1\le n\le\frac{p-1+m}{2}$ & $\frac{q^{((p+1)n-pm+\frac{p+1}{2})k}}{1+q^{(p+1)k+n-m}}-\frac{q^{((p+1)n-m+\frac{p+1}{2})k+\hlf m(2n-m+1)}}{1+q^{(p+1)k+n}}$ \\[3mm]
\hline
$\widetilde{E}_{n,m}$ & $1\le m\le p-2,\ m+1\le n\le\frac{p+m}{2}$ & $\frac{q^{((p+1)n-pm-\frac{p+1}{2})k}}{1+q^{(p+1)k+n-m}}-\frac{q^{((p+1)n-m-\frac{p+1}{2})k+\hlf m(2n-m-1)}}{1+q^{(p+1)k+n}}$ \\[3mm]
\hline
\end{tabular}}
\caption{Characters for representations of the lower series.  Here $p$, $a$, $b$, $n$, and $m$ are integers, and $x$ is a positive real number.}
\label{tab:LSChars}
\end{table}

\section{The algebras}
\label{sec:Algebras}

As pointed out in \cite{Figueroa-OFarrill:1996tnk}, the superconformal algebra which appears for an $\mathcal{N}=1$ superconformal sigma model whose target space has $\Spin(7)$ holonomy is just one example of a family of algebras labeled by the central charge $c$.  This family is denoted (in the conventions of \cite{Bouwknegt:1992wg}) $\SW(3/2,2)$.  Each member of the family includes the stress tensor $T(z)$ and the supersymmetry generator $G(z)$, with conformal weights $h=2$ and $h=3/2$ respectively, along with two more chiral operators\footnote{We mostly follow the conventions of \cite{Gepner:2001px}, but we have redefined the weight $5/2$ operator.  Our $M(z)$ relates to their $U(z)$ by
\be
M(z)=\frac{\sqrt{(15-c)(21+4c)}}{3(12+c)}U(z)+\frac{15-c}{6(12+c)}\p G(z).
\ee
Note that this means that for $c=12$, our $M(z)$ is not the same as the $M(z)$ used in \cite{Benjamin:2014kna}.  The relation is $M_{here}=M_{there}/8$, just as $A_{here}=X_{there}/8$.}, a bosonic operator $A(z)$ with conformal weight $h=2$ and a fermionic operator $M(z)$ with weight $h=5/2$.

Performing a standard mode expansion,
\be
T(z)=\sum_{m\in\Z}L_mz^{-m-2},\qquad G(z)=\sum_{r\in\Z+\n}G_rz^{-r-\tfrac{3}{2}},\non
\ee
\be
A(z)=\sum_{m\in\Z}A_mz^{-m-2},\qquad M(z)=\sum_{r\in\Z+\n}M_rz^{-r-\tfrac{5}{2}},
\ee
where $\n=0$ in the NS sector and $\n=\hlf$ in the R sector, then the mode algebra for these operators is given in appendix \ref{app:ModeAlgebra}.

\subsection{Classification}

From the commutation relation (\ref{eq:AACommutator}), we see that the $A_m$ modes generate a Virasoro sub-algebra with central charge\footnote{Note that $A(z)$ is not a Virasoro primary with respect to $T(z)$.  Rather, we can write $T(z)=A(z)+B(z)$, with $A$ and $B$ generating commuting Virasoro algebras with central charges $c_A$ and $c_B=c-c_A$.}
\be
c_A=\frac{c\lp 15-c\rp}{3\lp 12+c\rp}=1-\frac{\lp c-6\rp^2}{3\lp 12+c\rp}.
\ee

For a unitary theory, we must of course have $c\ge 0$, to ensure that none of the Virasoro descendants of the vacuum have negative norm.  Identical arguments show that we must also have $c_A\ge 0$, so that none of the $A$ descendants of the vacuum have negative norm.  

If $c=6$, then we have $c_A=1$.  It would be very interesting to analyze this case in more detail, but we leave that to future work.  Here we will focus on the alternative situation in which $c_A<1$, and hence if our theory is unitary then $A$ must generate a unitary Virasoro minimal model algebra.  This means that
\be
c_A=1-\frac{6}{p\lp p+1\rp},\qquad p=3,4,\cdots.
\ee
For instance $p=3$ corresponds to the critical Ising model, $p=4$ to the tri-critical Ising model, $p=5$ to the 3-state Potts model, and so on.  For each choice of $p$, there are then two distinct choices for $c$, either
\be
c_p^{(1)}=6+\frac{18}{p},\qquad c_p^{(2)}=6-\frac{18}{p+1}.
\ee
We will refer to the series of algebras with $c>6$ as the upper series, and with $c<6$ as the lower series.  

\subsection{Minimal model conventions}

The $A_m$ modes generate a subalgebra which is isomorphic to the Virasoro algebra for the $p$'th unitary minimal model.  Since $[A_0,L_0]=0$, we can label all of the states in our theory by their $L_0$ eigenvalue $h$ and their $A_0$ eigenvalue $a$, i.e.\ by their conformal weight and their minimal model weight.

The $p$'th unitary minimal model has only a finite number of lowest weight representations, with weights
\be
\label{eq:MMWeights}
h_{n,m}^{(p)}=\frac{\lp (p+1)n-pm\rp^2-1}{4p(p+1)},\qquad 1\le m\le n<p.
\ee

The characters for the associated minimal model representations are then given by
\be
\label{eq:MMChars}
\chi^{(p)}_{n,m}(q)=\frac{1}{\eta(q)}\ls\sum_{k\in\Z}q^{p(p+1)\lp k+\frac{(p+1)n-pm}{2p(p+1)}\rp^2}-\sum_{k\in\Z}q^{p(p+1)\lp k+\frac{(p+1)n+pm}{2p(p+1)}\rp^2}\rs,
\ee
where
\be
\eta(q)=q^{\frac{1}{24}}\prod_{n=1}^\infty\lp 1-q^n\rp
\ee
is the usual Dedekind eta function.

It will often be useful for us to use the formula above for integers $m$ and $n$ which do not lie in the given range.  In that case we have relations
\be
h^{(p)}_{n,m}=h^{(p)}_{n+pk,m+(p+1)k}=h^{(p)}_{-n,-m},
\ee
for arbitrary $k\in\Z$.  These relations also apply to the characters.

\section{The unitary representations}
\label{sec:Representations}

For all of these $\SW(3/2,2)$ algebras, the unitary representations were classified by Gepner and Noyvert~\cite{Gepner:2001px}.  We review their classification here, adding some extra notation so that we can refer to the representations through the rest of the paper.  Later, in section \ref{subsec:NewExpressions}, we will give an alternative description of the continuous representations which unifies the several different classes in each case below into a single family.

\subsection{Upper series}
\label{subsec:UpperSeriesReps}

\subsubsection{Continuous representations}
\label{subsubsec:UCont}

In the NS sector they found three classes of continuous representations (sometimes also called massive representations) for the upper series, which we will label X, Y, and Z.  In each of the cases listed below, $x$ can be any positive real number.  In the limit $x\rr 0$, the continuous representation will split into a sum of discrete representations according to threshold relations which we will elaborate in section~\ref{sec:ThresholdRelations}.

\begin{itemize}
\item In class X, we have $a(X^{(p)}_x)=0=h^{(p)}_{1,1}$ and $h(X^{(p)}_x)=x$.
\item In class Y, for each $2\le n\le p/2$, we have
\be
a(Y^{(p)}_{n;x})=\frac{\lp p(n-1)+2n-1\rp^2-1}{4p(p+1)}=h^{(p)}_{2n-1,n},\ h(Y^{(p)}_{n;x})=\frac{(n-1)\lp p(n-1)+2n\rp}{2p}+x.
\ee
\item In class Z, for each $1\le n\le p-2$ and $n+1\le m\le\tfrac{p+n}{2}$, we have
\be
a(Z^{(p)}_{n,m;x})=\frac{\lp pm-(p+1)n\rp^2-1}{4p(p+1)}=h^{(p)}_{p-n,p-m+1},
\ee
\be
h(Z^{(p)}_{n,m;x})=\frac{p(m-n)^2-(m-n)(2m-1)+m+1}{2p}+x.
\ee
\end{itemize}

In the R sector there were two more classes\footnote{This corrects a couple of typos in the corresponding table in Appendix B.2 of~\cite{Gepner:2001px}.}, V and W, and in each case there is a two-fold degeneracy in the lowest weight state in the representation.

\begin{itemize}
\item In class V, for each $n$ in the range $1\le n\le \tfrac{p-1}{2}$, the lowest weight states both have
\be
h(V^{(p)}_{n;x})=\frac{\lp 2n-1\rp\lp 2n+1-p\rp+2pn^2}{4p}+x,
\ee
and they have distinct values for $a$,
\be
a_+(V^{(p)}_{n;x})=h^{(p)}_{2n,n},\qquad a_-(V^{(p)}_{n;x})=h^{(p)}_{2n,n+1}.
\ee
\item In class W we have
\be
a_+(W^{(p)}_{n,m;x})=h^{(p)}_{p-n,p-m},\qquad a_-(W^{(p)}_{n,m;x})=h^{(p)}_{p-n,p+1-m},\non
\ee
\be
h(W^{(p)}_{n,m;x})=\frac{2p\lp m-n\rp^2-4m\lp m-n\rp+2p\lp m-n\rp+p+3}{4p}+x,
\ee
where the integers $n$ and $m$ satisfy either $1\le n\le p-3$, $n+1\le m\le \tfrac{p-1+n}{2}$ or $1\le n=m\le\tfrac{p}{2}$.  In the latter case the usual labeling convention for minimal model weights suggests that we should rewrite $a_-$ as $a_-(W^{(p)}_{n,n;x})=h^{(p)}_{n,n}$, leaving $a_+$ as $a_+(W^{(p)}_{n,n;x})=h^{(p)}_{p-n,p-n}$.  Note also that in the case of $p$ even and $n=m=\tfrac{p}{2}$, then we have $a_+=a_-$, and the two ground states have completely degenerate quantum numbers.
\end{itemize}

\subsubsection{Discrete representations}

Gepner and Noyvert also tell us the discrete representations of our algebras.  For the upper series, they tell us there are four classes of discrete NS representations:
\begin{itemize}
\item Class $A$ (the vacuum), has $a(A^{(p)})=h^{(p)}_{1,1}=0$, $h(A^{(p)})=0$.
\item Class $B$, labeled by $1\le n\le \tfrac{p}{2}-1$, with
\be
a(B^{(p)}_n)=h^{(p)}_{2n+1,n},\qquad h(B^{(p)}_n)=\frac{p(n+1)^2+2n(n+1)-2p}{2p}.
\ee
\item Class $C$, labeled by $1\le n\le p-2$ and $n+1\le m\le\tfrac{p+n}{2}$, with
\be
a(C^{(p)}_{n,m})=h^{(p)}_{p-n,p+1-m},\qquad h(C^{(p)}_{n,m})=\frac{p(m-n)^2-(m-n)(2m-1)+m+1}{2p}.
\ee
\item Class $D$, labeled by $2\le n\le p-1$ and $\tfrac{n}{2}\le m\le n-1$, with
\be
\label{eq:DQuantumNumbers}
a(D^{(p)}_{n,m})=h^{(p)}_{n,m},\qquad h(D^{(p)}_{n,m})=\frac{p(n-m)^2+(n-m)(2m+1)-m+1}{2p}.
\ee
\end{itemize}

In the R sector we have five more.  The first two classes, which are Ramond ground states with $h=c_p^{(1)}/24$, do not have the lowest weight degeneracy we have mentioned previously (since they are annihilated by both $G_0$ and $M_0$), while the last three do.
\begin{itemize}
\item Class $E$, with $a(E^{(p)})=\tfrac{p+3}{4p}=h^{(p)}_{2,1}$ and $h(E^{(p)})=a(E^{(p)})$.
\item Class $F$, labeled by $1\le n\le p-1$.  These all have $h(F^{(p)}_n)=\tfrac{p+3}{4p}$ (the same as the $E$ state) and
\be
a(F^{(p)}_n)=\frac{n^2-1}{4p\lp p+1\rp}=h^{(p)}_{n,n}.
\ee
\item Class $G$, labeled a pair of integers with $1\le n\le p-3$ and $n+1\le m\le \tfrac{p-1+n}{2}$.  There are two ground states which both have
\be
h(G^{(p)}_{n,m})=\frac{2p\lp m-n\rp^2-4m\lp m-n\rp+2p\lp m-n\rp+p+3}{4p},
\ee
and have distinct minimal model weights
\be
a_+(G^{(p)}_{n,m})=h^{(p)}_{p-n,p-m},\qquad a_-(G^{(p)}_{n,m})=h^{(p)}_{p-n,p+1-m}.
\ee
\item Class $H$, labeled by $1\le n\le p-3$, $n+2\le m\le \tfrac{p+1+n}{2}$ has
\be
a_+(H^{(p)}_{n,m})=h^{(p)}_{p-n,p-m},\qquad a_-(H^{(p)}_{n,m})=h^{(p)}_{p-n,p+1-m},\non
\ee
\be
h(H^{(p)}_{n,m})=\frac{2p\lp m-n\rp^2+4\lp m-1\rp\lp 1-m+n\rp+2p\lp m-n\rp-3\lp p-1\rp}{4p}.
\ee
\item Finally, there is a class $I$ of discrete states, labeled by $2\le n\le\tfrac{p-1}{2}$, which were missed by~\cite{Gepner:2001px} in their Appendix B.2.  The discrepancy only appears for $p\ge 5$.  There are three related arguments for why these states must be included.  First, we expect that there should be the same number of discrete states in both the NS and R sectors (this is already true for the continuous representations).  Second is the related observation that we have a spectral flow type operation coming from tensoring states with the class $E$ Ramond ground state.  Applying this to $B$ states produces the $I$ representations.  And finally, the threshold relations (discussed in more detail below, in section~\ref{sec:ThresholdRelations}) require the $I_n$ representations as well.  These states have
\be
a_+(I^{(p)}_n)=h^{(p)}_{2n,n-1},\qquad a_-(I^{(p)}_n)=h^{(p)}_{2n,n},\non
\ee
\be
h(I^{(p)}_n)=\frac{p(2n^2+2n-3)+4n^2-1}{4p}.
\ee
\end{itemize}

\subsection{Lower series}

\subsubsection{Continuous representations}

For $c=6-\tfrac{18}{p+1}$, we have only one NS sector class.
\begin{itemize}
\item Class $\widetilde{X}$ (the tilde is to distinguish it from the upper series classes), labeled by $1\le m\le p-2$ and $m\le n\le\tfrac{p-2+m}{2}$, with
\be
a(\widetilde{X}^{(p)}_{n,m;x})=\frac{\lp (p+1)n-pm\rp^2-1}{4p(p+1)}=h^{(p)}_{n,m},
\ee
\be
h(\widetilde{X}^{(p)}_{n,m;x})=\frac{p(n-m)^2+(3n-m-1)(n-m+1)-m}{2(p+1)}+x.
\ee
\end{itemize}

Similarly, in the R sector we have one class, 
\begin{itemize}
\item Class $\widetilde{V}$, with
\be
a_+(\widetilde{V}^{(p)}_{n,m;x})=h^{(p)}_{n,m},\qquad a_-(\widetilde{V}^{(p)}_{n,m;x})=h^{(p)}_{n-1,m},\non
\ee
\be
h(\widetilde{V}^{(p)}_{n,m;x})=\frac{2p\lp n-m\rp^2-2\lp n-m\rp\lp 1+m-3n\rp-2p\lp n-m\rp+p-2}{4\lp p+1\rp}+x,
\ee
where we either have $1\le m\le p-3$ and $m+1\le n\le\tfrac{p-1+m}{2}$ or we have $2\le n=m\le\tfrac{p+1}{2}$.  In the latter case we should rewrite $a_-$ as $a_-(\widetilde{V}^{(p)}_{n,n;x})=h^{(p)}_{p+1-n,p+1-n}$.  If $p$ is odd then there is one particular state, with $n=m=\tfrac{p+1}{2}$, for which $a_+=a_-$.
\end{itemize}

\subsubsection{Discrete representations}

In this case there are two classes of discrete representations in the NS sector,
\begin{itemize}
\item Class $\widetilde{A}$ labeled by $1\le m\le p-1$ and $m\le n\le\tfrac{p+m-1}{2}$, with
\be
a(\widetilde{A}^{(p)}_{n,m})=h^{(p)}_{n,m},\qquad h(\widetilde{A}^{(p)}_{n,m})=\frac{p\lp n-m\rp^2+\lp 3n-m-1\rp\lp n-m+1\rp-m}{2\lp p+1\rp}.
\ee
\item Class $\widetilde{B}$ labeled by $1\le m\le p-2$ and $m+1\le n\le\tfrac{p+m}{2}$, with
\be
a(\widetilde{B}^{(p)}_{n,m})=h^{(p)}_{n,m},\qquad h(\widetilde{B}^{(p)}_{n,m})=\frac{p\lp n-m\rp^2+\lp 3n-m+1\rp\lp n-m-1\rp+m}{2\lp p+1\rp}.
\ee
\end{itemize}

In the R sector we have three classes of representations.  The first, with $h=c_p^{(2)}/24$, is a nondegenerate ground state, while the other two have a pair of degenerate lowest weight states.
\begin{itemize}
\item Class $\widetilde{C}$, labeled by an integer $m$ with $1\le m\le p-1$ and
\be
a(\widetilde{C}^{(p)}_m)=h^{(p)}_{m,m},\qquad h(\widetilde{C}^{(p)}_m)=\frac{p-2}{4\lp p+1\rp}.
\ee
\item Class $\widetilde{D}$ labeled by $1\le m\le p-3$ and $m+1\le n\le\tfrac{p-1+m}{2}$ with
\be
a_+(\widetilde{D}^{(p)}_{n,m})=h^{(p)}_{n+1,m},\qquad a_-(\widetilde{D}^{(p)}_{n,m})=h^{(p)}_{n,m},
\ee
and
\be
h(\widetilde{D}^{(p)}_{n,m})=\frac{p\lp n-m\rp^2+\lp n-m\rp\lp 1-m+3n\rp+p\lp n-m\rp+\frac{p}{2}-1}{2\lp p+1\rp}.
\ee
\item Class $\widetilde{E}$ labeled by $1\le m\le p-2$ and $m+1\le n\le\tfrac{p+m}{2}$ with
\be
a_+(\widetilde{E}^{(p)}_{n,m})=h^{(p)}_{n,m},\qquad a_-(\widetilde{E}^{(p)}_{n,m})=h^{(p)}_{n-1,m},
\ee
and
\be
h(\widetilde{E}^{(p)}_{n,m})=\frac{p\lp n-m\rp^2-\lp n-m\rp\lp 1+m-3n\rp-p\lp n-m\rp+\frac{p}{2}-1}{2\lp p+1\rp}.
\ee
\end{itemize}

\section{Character preliminaries}
\label{sec:Embedding}

\subsection{Kac determinant results}

In the process of their classification, Gepner and Noyvert~\cite{Gepner:2001px} analyzed the Kac determinant for these theories.  They found three classes of potential singular (and hence null) descendants for a given lowest weight state, labeled $f_{m,n}$, $g_{j,k}$, or $d_\ell$.  Gepner and Noyvert give the conditions for the existence of these states in terms of $h$ and $a$ of the lowest weight states, but we will find it more convenient to introduce new quantities $\Phi$ and $\Lambda$, in terms of which the conditions are more easily stated.  We will use shorthand $|\Phi,\Lambda\rangle$ for a state with the given quantum numbers, and we'll present the results for each of the two series, and for each sector (NS and R).

In the following sections, we will construct embedding diagrams in the $\Phi\Lambda$-plane using the results from the descendants, and by making some assumptions about multiplicities, we will deduce characters for each type of representation.

\subsubsection{Upper series, NS sector}

Consider the upper series, with $c=c_p^{(1)}=6+18/p$.  We first redefine the quantum numbers of any state by
\be
\label{eq:PhiDef}
\Phi=2\sqrt{4p\lp p+1\rp a+1},\qquad \Lambda=\sqrt{\lp p+4\rp^2+16p\lp p+1\rp a-8p\lp p+2\rp h}.
\ee
The quantity under the square root in the expression for $\Lambda$ can be negative, but only for continuous representations with sufficiently large $h$, and in those cases $\Lambda$ won't play a role, so we can assume the quantity is positive.  We will always take the positive branch of the square roots so that both $\Phi$ and $\Lambda$ are assumed to be positive.  Note that for a primary field with $a=h^{(p)}_{n,m}$, then $\Phi=2|(p+1)n-pm|\in 2\Z$.  Also, the inverse relations are
\be
a=\frac{\Phi^2-4}{16p\lp p+1\rp},\qquad h=\frac{\Phi^2-\Lambda^2}{8p\lp p+2\rp}+\frac{p+6}{8p}.
\ee
Contours of constant $h$ are hyperbolas in the $\Phi\Lambda$-plane.

Now, the results of~\cite{Gepner:2001px} can be summarized as follows.  For each pair of integers $(n,m)$ satisfying
\be
\lp p+2\rp m-pn=\Phi,\quad\mathrm{or}\quad\lp p+2\rp m-pn=-\Phi,\quad m,n>0,\ m+n\in 2\Z,
\ee
the state $|\Phi,\Lambda\rangle$ has a level $mn/2$ singular descendant, called an $f$ descendant and labeled by $f_{n,m}$ acting on the original state, with
\be
f_{n,m}\left|\Phi,\Lambda\right\rangle=\left|\Phi',\Lambda\right\rangle,
\ee
where the descendant has the same $\La$ as the parent state, and a new value
\be
\Phi'=\lp p+2\rp m+pn.
\ee

Similarly, a $g_{j,k}$ descendant (at level $jk$) corresponds to a solution of
\be
\Lambda=2\lp p+2\rp j+pk,\qquad j,k>0,
\ee
(so in particular we at least need $\Lambda\ge 3p+4$), and has
\be
g_{j,k}\left|\Phi,\Lambda\right\rangle=\left|\Phi,\Lambda'\right\rangle,
\ee
where $\Phi$ is unchanged and
\be
\Lambda'=\left|2\lp p+2\rp j-pk\right|.
\ee

Finally, a $d_\ell$ descendant happens each time we have a choice of signs $\eta_1$ and $\eta_2$ such that
\be
\ell=\frac{\eta_1\Phi+\eta_2\Lambda}{p+2},
\ee
is a positive odd integer.  Then
\be
d_\ell\left|\Phi,\Lambda\right\rangle=\left|\Phi',\Lambda'\right\rangle,
\ee
with
\be
\Phi'=\left|\Phi+2\eta_1p\right|,\qquad\Lambda'=\left|\Lambda-2\eta_2p\right|.
\ee
This descendant is level $\ell/2$.

In terms of a diagram in the $\Phi\Lambda$-plane, an $f$-descendant always lies to the right of its parent, a $g$-descendant always lies below its parent, while a $d$-descendants lie along diagonal lines\footnote{Note that because we take absolute values to restrict to the upper-right quadrant of the $\Phi\La$-plane, the lines can ``reflect'' off the axes.  See Figure~\ref{fig:ContinuousASpecial} for example.} with slope $\pm 1$.  For slope $-1$, the descendants always lie down and to the right, while for slope $+1$ they can lie either lie up and to the right, if the source state has $\Phi>\Lambda$, or down and to the left, if the source state has $\Phi<\Lambda$ (so that in both cases the weight $h$ increases).

\subsubsection{Upper series, R sector}

By inspecting the Ramond sector representations in section \ref{subsec:UpperSeriesReps}, one sees that each representation either has a single lowest weight state with a given $a$ and $h=(p+3)/4p$, or there are a pair of lowest weight states with the same $h>(p+3)/4p$ but (possibly) different $a_+$ and $a_-$, with $a_+\ge a_-$.  In the latter case, we will define
\be
\Phi=4\lp p+1\rp\lp a_+-a_-\rp,\quad\Lambda=\sqrt{8p\lp p+1\rp\lp a_++a_-\rp-8p\lp p+2\rp h+p^2+10p+16},
\ee
and then the inverse relations are (note that $a_+$ and $a_-$ are not independent)
\be
a_\pm=\frac{\lp\Phi\pm p\rp^2-4}{16p\lp p+1\rp},\qquad h=\frac{\Phi^2-\La^2}{8p\lp p+2\rp}+\frac{p+3}{4p}.
\ee
In this case we can think of the two states as forming a two-component vector
\be
\left|\Phi,\La\right\rangle=\lp\begin{matrix} a_+, h \\ a_-, h \end{matrix}\rp.
\ee

The classification of singular descendants is very similar to the NS sector, with a couple of minor differences.  Given any positive integers $m$ and $n$, with $m+n\in 2\Z+1$ and satisfying
\be
\left|\lp p+2\rp m-pn\right|=\Phi,
\ee
we have a level $mn/2$ descendant
\be
f_{n,m}\left|\Phi,\La\right\rangle=\left|\Phi',\La\right\rangle,\qquad\Phi'=\lp p+2\rp m+pn.
\ee

Given any positive integers $j$ and $k$ with
\be
2\lp p+2\rp j+pk=\La,
\ee
we have a level $jk$ descendant
\be
g_{j,k}\left|\Phi,\La\right\rangle=\left|\Phi,\La'\right\rangle,\qquad\La'=\left|2\lp p+2\rp j-pk\right|.
\ee

And finally, if there is a choice of signs $\eta_1$ and $\eta_2$ such that
\be
\ell=\frac{\eta_1\Phi+\eta_2\La}{p+2}
\ee
is a positive even integer, then there is a level $\ell/2$ descendant
\be
d_\ell\left|\Phi,\La\right\rangle=\left|\Phi',\La'\right\rangle,\qquad\Phi'=\left|\Phi+2\eta_1p\right|,\quad\La'=\left|\La-2\eta_2p\right|.
\ee

For the degenerate cases with $h=(p+3)/4p$, there is only one ground state.  Here we have two choices for the corresponding $\Phi$ and $\La$,
\be
\Phi=\left|\pm p+2\sqrt{1+4p\lp p+1\rp a}\right|,\qquad\La=\Phi,
\ee
so
\be
a=\frac{\lp\Phi\mp p\rp^2-4}{16p\lp p+1\rp},\qquad h=\frac{p+3}{4p}.
\ee
These states always have $d_0$ descendants, removing the other state in the doublet from the representation and the character.  The fact that there are two options may seem like an ambiguity, but one can check that making either choice leads to a set of states in the representation with exactly the same $a$ and $h$ quantum numbers.

\subsubsection{Lower series, NS sector}

The situation is very similar to the upper series.  We define $\Phi$ as before in (\ref{eq:PhiDef}), while $\Lambda$ gets modified to
\be
\wtL=\sqrt{\lp p-3\rp^2+16p\lp p+1\rp a-8\lp p-1\rp\lp p+1\rp h}.
\ee

For each pair of integers $n$ and $m$ satisfying
\be
\lp p+1\rp n-\lp p-1\rp m=\pm\Phi,\quad m,n>0, m+n\in 2\Z,
\ee
we have an $f$ descendant
\be
f_{n,m}\left|\Phi,\wtL\right\rangle=\left|\Phi',\wtL\right\rangle,\qquad\Phi'=\lp p+1\rp n+\lp p-1\rp m.
\ee

For each $j$ and $k$ satisfying
\be
2\lp p-1\rp j+\lp p+1\rp k=\wtL,\quad j,k>0,
\ee
we have a $g$ descendant
\be
g_{j,k}\left|\Phi,\wtL\right\rangle=\left|\Phi,\wtL'\right\rangle,\qquad\wtL'=\left|2\lp p-1\rp j-\lp p+1\rp k\right|.
\ee

And lastly, there is a $d$ descendant if we can find a choice of signs $\eta_1$ and $\eta_2$, and a positive odd integer $\ell$, such that
\be
\ell=\frac{\eta_1\Phi+\eta_2\wtL}{p-1},
\ee
in which case
\be
d_\ell\left|\Phi,\wtL\right\rangle=\left|\Phi',\wtL'\right\rangle,\qquad\Phi'=\left|\Phi+2\eta_1\lp p+1\rp\right|,\quad\wtL'=\left|\wtL-2\eta_2\lp p+1\rp\right|.
\ee

\subsubsection{Lower series, R sector}

Briefly, for the lower series R sector, each representation has a single lowest weight state if $h=(p-2)/4(p+1)$ or a pair of degenerate lowest weight states if $h>(p-2)/4(p+1)$.  In the latter case we define
\be
\Phi=4p\lp a_+-a_-\rp,\qquad\wtL=\sqrt{8p\lp p+1\rp\lp a_++a_-\rp-8\lp p-1\rp\lp p+1\rp h+p^2-8p+7},
\ee
or
\be
a_\pm=\frac{\lp\Phi\pm\lp p+1\rp\rp^2-4}{16p\lp p+1\rp},\qquad h=\frac{\Phi^2-\wtL^2}{8\lp p-1\rp\lp p+1\rp}+\frac{p-2}{4\lp p+1\rp}.
\ee
In this case the states form a two-component vector
\be
\left|\Phi,\wtL\right\rangle=\lp\begin{matrix} a_+,h \\ a_-,h \end{matrix}\rp.
\ee

For these representations, given any positive integers  $m$ and $n$ with $m+n\in 2\Z+1$ and satisfying
\be
\left|\lp p-1\rp m-\lp p+1\rp n\right|=\Phi,
\ee
we have a level $nm/2$ descendant
\be
f_{n,m}\left|\Phi,\wtL\right\rangle=\left|\Phi',\wtL\right\rangle,\qquad\Phi'=\lp p-1\rp m+\lp p+1\rp n.
\ee

Given positive integers $j$ and $k$ obeying
\be
2\lp p-1\rp j+\lp p+1\rp k=\wtL,
\ee
there is a level $jk$ descendant
\be
g_{j,k}\left|\Phi,\wtL\right\rangle=\left|\Phi,\wtL'\right\rangle,\qquad\wtL'=\left|2\lp p-1\rp j-\lp p+1\rp k\right|.
\ee

And then there is a level $\ell/2$ descendant $d_\ell$ provided there is a choice of signs $\eta_1$ and $\eta_2$ so that
\be
\ell=\frac{\eta_1\Phi+\eta_2\wtL}{p-1}
\ee
is positive and even, in which case
\be
d_\ell\left|\Phi,\wtL\right\rangle=\left|\Phi',\wtL'\right\rangle,\qquad\Phi'=\left|\Phi+2\eta_1\lp p+1\rp\right|,\quad\wtL'=\left|\wtL-2\eta_2\lp p+1\rp\right|.
\ee

For the degenerate case $h=(p-2)/4(p+1)$, there is only one ground state.  We can take
\be
\Phi=p+1+2\sqrt{1+4p\lp p+1\rp a},\qquad\wtL=\Phi,
\ee
and proceed as before.  These states always have $d_0$ descendants.

\section{Continuous representation characters}
\label{sec:ContinuousRepCharacters}

\subsection{Embedding diagrams}

We'll run through our procedure for the NS sector of the upper series in some detail.  For the other cases we will be more terse.

As reviewed in section \ref{subsubsec:UCont}, three classes of continuous representations were found.  In class X, there is a unique possibility that has $a(X_x)=0$, and hence $\Phi(X_x)=2$, while $h(X_x)=x>0$ so $\Lambda(X_x)=\sqrt{(p+4)^2-8p(p+2)x}$.  We write $|X_x\rangle=|2,\Lambda(X_x)\rangle$, or simply $|2,\Lambda\rangle$.  Searching first for $f$ descendants, we look for solutions to the equations
\be
\lp p+2\rp m-pn=\pm 2,\qquad n,m>0,\quad n+m\in 2\Z.
\ee
For the plus sign, the most general solution is $n=1+(p+2)k$, $m=1+pk$, where $k\ge 0$.  For the minus sign the solutions are $n=p+1+(p+2)k$, $m=p-1+pk$, where again $k\ge 0$.  There are no $g$ descendants, since $\Lambda$ is too small (when it is even real).  In order to have any $d$ descendants for $x>0$, we would need $\Lambda(X_x)=p$, which happens only at the isolated point $x=1/p$.  We'll treat this special case below and continue for now with the case of generic $x$.

What about descendants of singular vectors?  The $f_{n,m}$ descendants constructed above will be
\bea
f_{1+(p+2)k,1+pk}\left|2,\Lambda\right\rangle &=& \left|2p+2+2p(p+2)k,\Lambda\right\rangle,\\
f_{p+1+(p+2)k,p-1+pk}\left|2,\Lambda\right\rangle &=& \left|2p^2+2p-2+2p(p+2)k,\Lambda\right\rangle.
\eea
Since these have the same $\Lambda$ value, they of course still have no $g$ descendants, and one can check that for generic $x$ (i.e.\ $x>0$, $x\ne 1/p$) there are no $d$ descendants either.  They do have further $f_{n,m}$ descendants, however.  For $|2p+2+2p(p+2)k,\Lambda\rangle$ we have
\bea
f_{p+1+(p+2)j,p+1+p(2k+j)}\left|2p+2+2p(p+2)k,\Lambda\right\rangle &=& \left|2+2p(p+2)(k+j+1),\Lambda\right\rangle,\\
f_{p+3+(p+2)(2k+j),p-1+pj}\left|2p+2+2p(p+2)k,\Lambda\right\rangle &=& \left|2p^2+4p-2+2p(p+2)(k+j),\Lambda\right\rangle,\non\\
\eea
where $j\ge 0$, while for $|2p^2+2p-2+2p(p+2)k,\Lambda\rangle$ we have
\begin{multline}
f_{1+(p+2)j,2p-1+p(2k+j)}\left|2p^2+2p-2+2p(p+2)k,\Lambda\right\rangle\\
=\left|2p^2+4p-2+2p(p+2)(k+j),\Lambda\right\rangle,
\end{multline}
\be
f_{2p+3+(p+2)(2k+j),1+pj}\left|2p^2+2p-2+2p(p+2)k,\Lambda\right\rangle=\left|2+2p(p+2)(k+j+1),\Lambda\right\rangle,
\ee
where again $j\ge 0$.  In both cases the products fall into the same sequences.  We have states of the form $|2+2p(p+2)k,\Lambda\rangle$ (which includes our original state $|2,\Lambda\rangle$) which again have no $g$ or $d$ descendants and have ($j\ge 0$)
\bea
f_{1+(p+2)j,1+p(2k+j)}\left|2+2p(p+2)k,\Lambda\right\rangle &=& \left|2p+2+2p(p+2)(k+j),\Lambda\right\rangle,\\
f_{p+1+(p+2)(2k+j),p-1+pj}\left|2+2p(p+2)k,\Lambda\right\rangle &=& \left|2p^2+2p-2+2p(p+2)(k+j),\Lambda\right\rangle,\non\\
\eea
and states $|2p^2+4p-2+2p(p+2)k,\Lambda\rangle$ with
\begin{multline}
f_{p+1+(p+2)j,3p-1+p(2k+j)}\left|2p^2+4p-2+2p(p+2)k,\Lambda\right\rangle\\
=\left|2p^2+2p-2+2p(p+2)(k+j+1),\Lambda\right\rangle,
\end{multline}
\begin{multline}
f_{2p+5+(p+2)(2k+j),1+pj}\left|2p^2+4p-2+2p(p+2)k,\Lambda\right\rangle\\
=\left|2p+2+2p(p+2)(k+j+1),\Lambda\right\rangle.
\end{multline}
At this point we are not introducing any new states.  

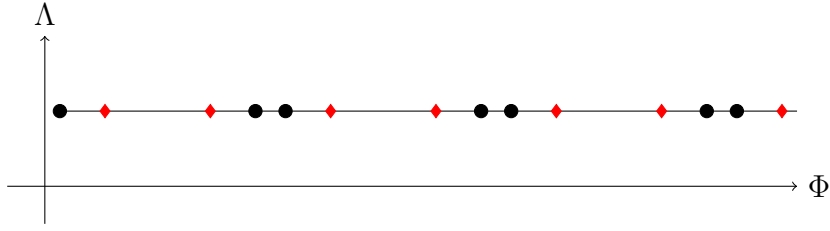
\begin{figure}
\label{fig:ContinuousA}
\centering
\begin{tikzpicture}[scale=0.5]
   \pgfsetplotmarksize{2.5pt}
   \draw[->] (-1,0) -- (20,0) node[right] {$\Phi$};
   \draw[->] (0,-1) -- (0,4) node[above] {$\Lambda$};
   \draw (0.4,2) -- (20,2);
   \foreach \k in {0,...,3} {
      \node at (6*\k+0.4,2) {\pgfuseplotmark{*}};
      \node[color=red] at (6*\k+1.6,2) {\pgfuseplotmark{diamond*}};
   }
   \foreach \k in {0,...,2} {
      \node[color=red] at (6*\k+4.4,2) {\pgfuseplotmark{diamond*}};
      \node at (6*\k+5.6,2) {\pgfuseplotmark{*}};
   }
\end{tikzpicture}
\caption{The $\Phi\Lambda$ embedding diagram for the $p=3$ $X$ series continuous representation.  Black nodes contribute to the character with a plus sign, red nodes with a negative sign.}
\label{fig:ContinuousA}
\end{figure}

The structure of descendants can be summarized in a diagram, as shown in Figure \ref{fig:ContinuousA}.  The black circle on the left is the original primary state $|2,\Lambda\rangle$.  All of its descendants, $|2p+2+2p(p+2)k,\Lambda\rangle$ and $|2p^2+2p-2+2p(p+2)k,\Lambda\rangle$, are represented as red diamonds.  The other black circles represent the states $|2+2p(p+2)k,\Lambda\rangle$ and $|2p^2+4p-2+2p(p+2)k\rangle$.  Each red diamond is an $f$ descendant of every black circle to its left, and has every black circle to its right as a descendant.  Similarly, each black circle is a descendant of every red diamond to its left, and has every red diamond to its right as a descendant.

Our notation here glosses over an important subtlety.  We are treating each $|\Phi,\Lambda\rangle$ as a unique state.  In fact, it can happen that there are several states with the same quantum numbers $\Phi$ and $\Lambda$.  We currently have no definitive way of evaluating this possibility in general.  At very low levels one can explicitly compute the singular descendants and compare, but this quickly becomes impractical, even by computer.  Instead, in this paper we shall make assumptions about the multiplicities and overlaps (i.e.\ when a given set of quantum numbers are obtained as a descendant in different ways), and then run as many checks as possible (e.g.\ all coefficients are nonnegative integers, threshold relations are obeyed) to confirm our hypotheses.  Obviously, it would be desirable to fill in these gaps.

From the embedding diagram (and the assumptions about multiplicities) we can construct the characters.  For each state we need to include all of its possible descendants built using $L_{-n}$, $A_{-n}$, $G_{-r}$, and $M_{-r}$ oscillators.  This means that to a state $|\Phi,\Lambda\rangle$ we associate a contribution $P_{NS}(\tau)q^{h-\tfrac{c}{24}}$, where
\be
\label{eq:PNS}
P_{NS}(\tau)=\prod_{n=1}^\infty\frac{\lp 1+q^{n-\hlf}\rp^2}{\lp 1-q^n\rp^2},\qquad h=\frac{\Phi^2-\Lambda^2+\lp p+6\rp\lp p+2\rp}{8p\lp p+2\rp}.
\ee
To build the character, we start with this contribution for the original state.  Then, moving to the right in the diagram, we know that each node represents a singular vector; a null state whose contribution needs to be removed.  So for the first two diamonds, since they have each been added as part of the descendants of the primary, we must subtract their contribution (to signify this we have colored them red in the diagram).  For the next pair of black dots, they have three antecedents - the primary and each of the two red diamonds (we assume that the same state is a descendant of each of the red diamond states, i.e.\ that the multiplicity of the sub-singular state with these quantum numbers is one).  Thus they have been added once and subtracted twice, and to get them to zero we must add them back in.  We color them black to indicate this.  Similarly, each of the next pair of diamonds has been added three times and subtracted twice, and so we must subtract them.  Proceeding in this way, we build the entire character,
\bea
\label{eq:ContinuousXCharacter}
\chi[X^{(p)}_x](q) &=& P_{NS}(\tau)q^{x-\tfrac{1}{4}-\tfrac{3}{4p}}\sum_{k=0}^\infty q^{\hlf p(p+2)k^2}\ls\vphantom{q^{(p^2+2p-1)k+\hlf(p^2+2p-2)}} q^k-q^{(p+1)k+\hlf}\right.\non\\
&& \qquad\left. -q^{(p^2+p-1)k+\hlf(p^2-1)}+q^{(p^2+2p-1)k+\hlf(p^2+2p-2)}\rs.
\eea
This result matches the $p=3$ result of \cite{Benjamin:2014kna} and also, by comparing $q$-expansions to very high order, the result of \cite{Eguchi:2003yy}.

The expression can be compressed a little bit by rewriting the sum to run over all integers instead of non-negative integers,
\be
\chi[X^{(p)}_x](q)=P_{NS}(\tau)q^{x-\tfrac{1}{4}-\tfrac{3}{4p}}\sum_{k\in\Z}^\infty q^{\hlf p(p+2)k^2}\ls q^k-q^{(p+1)k+\hlf}\rs.
\ee

Similar calculations work for the other classes of continuous representations.  For the $Y$ series we find states
\be
\left|2(n-1)p+4n-2+2p(p+2)k,\Lambda\right\rangle,\quad\mathrm{and}\quad\left|2p^2-2(n-3)p-4n+2+2p(p+2)k,\Lambda\right\rangle,
\ee
that contribute with a positive sign, and
\be
\left|2np+4n-2+2p(p+2)k,\Lambda\right\rangle\quad\mathrm{and}\quad\left|2p^2-2(n-2)p-4n+2+2p(p+2)k,\Lambda\right\rangle,
\ee
which get a negative sign, leading to
\begin{align}
\chi[Y^{(p)}_{n;x}](q)=\ & P_{NS}(\tau)q^{x+\tfrac{(n-1)(p(n-1)+2n)}{2p}-\tfrac{1}{4}-\tfrac{3}{4p}}\sum_{k=0}^\infty q^{\hlf p(p+2)k^2}\ls\vphantom{q^{(p^2-(n-3)p-2n+1)k+\hlf(p^2-2(n-2)p-4n+2)}} q^{((n-1)p+2n-1)k}\right.\non\\
& \qquad\left. -q^{(np+2n-1)k+\hlf(2n-1)}-q^{(p^2-(n-2)p-2n+1)k+\hlf(p^2-2(n-1)p-2n+1)}\right.\non\\
& \qquad\left. +q^{(p^2-(n-3)p-2n+1)k+\hlf(p^2-2(n-2)p-4n+2)}\rs\non\\
=\ & P_{NS}(\tau)q^{x+\tfrac{(n-1)(p(n-1)+2n)}{2p}-\tfrac{1}{4}-\tfrac{3}{4p}}\non\\
& \quad\times\sum_{k\in\Z}^\infty q^{\hlf p(p+2)k^2}\ls q^{((n-1)p+2n-1)k}-q^{(np+2n-1)k+\hlf(2n-1)}\rs.
\end{align}
And for the $Z$ series,
\be
\left|2pm-2(p+1)n+2p(p+2)k,\Lambda\right\rangle,\quad\mathrm{and}\quad\left|2p^2+4p-2pm+2(p+1)n+2p(p+2)k,\Lambda\right\rangle,
\ee
with a positive sign and
\be
\left|2pm+2n+2p(p+2)k,\Lambda\right\rangle,\quad\mathrm{and}\quad\left|2p^2+4p-2pm-2n+2p(p+2)k,\Lambda\right\rangle,
\ee
with a negative sign, leading to characters
\begin{align}
\chi[Z^{(p)}_{n,m;x}](q)=\ & P_{NS}(\tau)q^{x+\tfrac{p(m-n)^2-(m-n)(2m-1)+m+1}{2p}-\tfrac{1}{4}-\tfrac{3}{4p}}\sum_{k=0}^\infty q^{\hlf p(p+2)k^2}\ls\vphantom{q^{(p^2+2p-pm+(p+1)n)k+\hlf(p^2+2p-2pm+2(p+1)n)}} q^{(pm-(p+1)n)k}\right.\non\\
& \qquad\left. -q^{(pm+n)k+\hlf n(2m-n)}-q^{(p^2+2p-pm-n)k+\hlf(p-n)(p+2-2m+n)}\right.\non\\
& \qquad\left. +q^{(p^2+2p-pm+(p+1)n)k+\hlf(p^2+2p-2pm+2(p+1)n)}\rs\non\\
=\ & P_{NS}(\tau)q^{x+\tfrac{p(m-n)^2-(m-n)(2m-1)+m+1}{2p}-\tfrac{1}{4}-\tfrac{3}{4p}}\non\\
& \quad\times\sum_{k\in\Z}^\infty q^{\hlf p(p+2)k^2}\ls q^{(pm-(p+1)n)k}-q^{(pm+n)k+\hlf n(2m-n)}\rs.
\end{align}
Again these agree perfectly with the results from the literature for $p=3$.

Similarly for the Ramond sector continuous representations we find
\begin{align}
\chi[V^{(p)}_{n;x}](q)=\ & P_R(\tau)q^{x+\frac{(2n-1)(2n+1-p)+2pn^2}{4p}-\frac{1}{4}-\frac{3}{4p}}\sum_{k=0}^\infty q^{\hlf p(p+2)k^2}\ls q^{\lp (p+2)n-\frac{p}{2}\rp k}\right.\non\\
& \qquad\left. -q^{\lp (p+2)n+\frac{p}{2}\rp k+n}-q^{\lp p^2+\frac{3p}{2}-(p+2)n\rp k+\hlf(p+1)(p-2n)}\right.\non\\
& \qquad\left. +q^{\lp p^2+\frac{5p}{2}-(p+2)n\rp k+\hlf(p+3)(p-2n)+n}\rs\non\\
=\ & P_R(\tau)q^{x+\frac{(2n-1)(2n+1-p)+2pn^2}{4p}-\frac{1}{4}-\frac{3}{4p}}\non\\
& \quad\times\sum_{k\in\Z}^\infty q^{\hlf p(p+2)k^2}\ls q^{\lp (p+2)n-\frac{p}{2}\rp k}-q^{\lp (p+2)n+\frac{p}{2}\rp k+n}\rs,\\
\chi[W^{(p)}_{n,m;x}](q) =\ & P_R(\tau)q^{x+\frac{2p(m-n)^2-4m(m-n)+2p(m-n)+p+3}{4p}-\frac{1}{4}-\frac{3}{4p}}\sum_{k=0}^\infty q^{\hlf p(p+2)k^2}\ls q^{\lp pm-(p+1)n+\frac{p}{2}\rp k}\right.\non\\
& \qquad\left. -q^{\lp p^2+\frac{3p}{2}-n-pm\rp k+\hlf(p-n)(p+1+n-2m)}-q^{\lp\frac{p}{2}+n+pm\rp k+\hlf n(2m-n+1)}\right.\non\\
& \qquad\left. +q^{\lp p^2+\frac{3p}{2}+(p+1)n-pm\rp k+\hlf n(2p+3-2m+n)+\hlf(p-n)(p+4n-2m)}\rs\non\\
=\ & P_R(\tau)q^{x+\frac{2p(m-n)^2-4m(m-n)+2p(m-n)+p+3}{4p}-\frac{1}{4}-\frac{3}{4p}}\non\\
& \quad\times\sum_{k\in\Z}^\infty q^{\hlf p(p+2)k^2}\ls q^{\lp pm-(p+1)n+\frac{p}{2}\rp k}-q^{\lp pm+n+\frac{p}{2}\rp k+\hlf n(2m-n+1)}\rs.\end{align}
where
\be
\label{eq:PR}
P_R(\tau)=2\prod_{n=1}^\infty\frac{\lp 1+q^n\rp^2}{\lp 1-q^n\rp^2}.
\ee

\subsection{Special case of $x=1/p$}

Let's revisit the $X$ representation character at the special value of $x=1/p$, so the lowest weight state is $|\Phi,\Lambda(X_{1/p})\rangle=|2,p\rangle$.  In this case the primary has a $d$-descendant,
\be
d_1\left|2,p\right\rangle=\left|2p+2,p\right\rangle.
\ee
There is also an $f$-descendant, $f_{1,1}|2,p\rangle$, but it is not distinct from the $d$-descendant.  Indeed, one can do an explicit computation in this case and show that there is only one singular vector at level-$1/2$, and so both descendants must coincide with it,
\be
d_1\left|2,p\right\rangle=f_{1,1}\left|2,p\right\rangle=\mathcal{N}\ls\lp p-2\rp G_{-1/2}+4\lp p+1\rp M_{-1/2}\rs\left|2,p\right\rangle=\left|2p+2,p\right\rangle,
\ee
where $\mathcal{N}$ is a normalization constant.

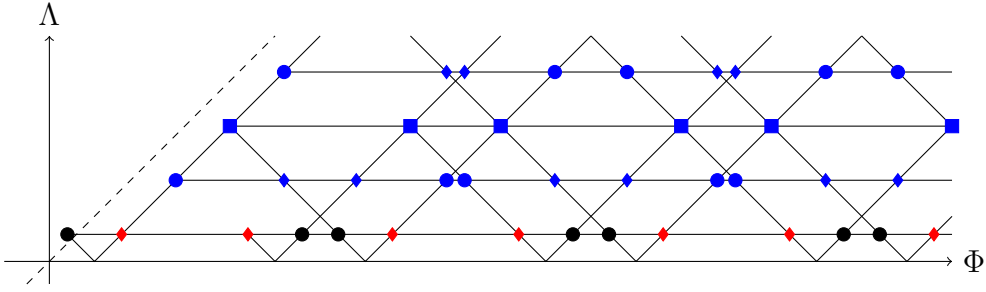
\begin{figure}
\centering
\begin{tikzpicture}[xscale=0.6,yscale=0.3]
   \pgfsetplotmarksize{2.5pt}
   \draw[->] (-1,0) -- (20,0) node[right] {$\Phi$};
   \draw[->] (0,-1) -- (0,10) node[above] {$\Lambda$};
   \draw[dashed] (-0.5,-1) -- (5,10);
   \draw (0.4,1.2) -- (20,1.2);
   \draw (0.4,1.2) -- (1,0) -- (6,10);
   \draw (2.8,3.6) -- (20,3.6);
   \draw (4,6) -- (20,6);
   \draw (5.2,8.4) -- (20,8.4);
   \draw (4.4,1.2) -- (5,0) -- (10,10);
   \draw (4,6) -- (7,0) -- (12,10);
   \draw (8,6) -- (11,0) -- (16,10);
   \draw (8,10) -- (13,0) -- (18,10);
   \draw (12,10) -- (17,0) -- (20,6);
   \draw (14,10) -- (19,0) -- (20,2);
   \draw (18,10) -- (20,6);
   \foreach \k in {0,...,3} {
      \node at (6*\k+0.4,1.2) {\pgfuseplotmark{*}};
      \node[color=red] at (6*\k+1.6,1.2) {\pgfuseplotmark{diamond*}};
   }
   \foreach \k in {0,...,2} {
      \node[color=red] at (6*\k+4.4,1.2) {\pgfuseplotmark{diamond*}};
      \node at (6*\k+5.6,1.2) {\pgfuseplotmark{*}};
   }
   \foreach \k in {0,...,2} {
      \node[color=blue] at (6*\k+2.8,3.6) {\pgfuseplotmark{*}};
      \node[color=blue] at (6*\k+5.2,3.6) {\pgfuseplotmark{diamond*}};
      \node[color=blue] at (6*\k+6.8,3.6) {\pgfuseplotmark{diamond*}};
   }
   \node[color=blue] at (9.2,3.6) {\pgfuseplotmark{*}};
   \node[color=blue] at (15.2,3.6) {\pgfuseplotmark{*}};
   \foreach \k in {0,...,2} {
      \node[color=blue] at (6*\k+4,6) {\pgfuseplotmark{square*}};
      \node[color=blue] at (6*\k+8,6) {\pgfuseplotmark{square*}};
   }
   \node[color=blue] at (5.2,8.4) {\pgfuseplotmark{*}};
   \node[color=blue] at (11.2,8.4) {\pgfuseplotmark{*}};
   \node[color=blue] at (17.2,8.4) {\pgfuseplotmark{*}};
   \node[color=blue] at (12.8,8.4) {\pgfuseplotmark{*}};
   \node[color=blue] at (18.8,8.4) {\pgfuseplotmark{*}};
   \node[color=blue] at (8.8,8.4) {\pgfuseplotmark{diamond*}};
   \node[color=blue] at (14.8,8.4) {\pgfuseplotmark{diamond*}};
   \node[color=blue] at (9.2,8.4) {\pgfuseplotmark{diamond*}};
   \node[color=blue] at (15.2,8.4) {\pgfuseplotmark{diamond*}};   
\end{tikzpicture}
\caption{The $\Phi\Lambda$ embedding diagram for the $p=3$ $X$ series continuous representation at the special value $x=1/3$.  Black nodes contribute to the character with a plus sign, red nodes with a negative sign, while blue nodes do not contribute to the character.}
\label{fig:ContinuousASpecial}
\end{figure}

In fact, all of the $f$-descendants themselves have $d$-descendants,
\bea
d_{1+2pk}\left|2+2p(p+2)k,p\right\rangle &=& \left|2p+2+2p(p+2)k,p\right\rangle,\\
d_{1+2pk}\left|2p+2+2p(p+2)k,p\right\rangle &=& \left|4p+2+2p(p+2)k,3p\right\rangle,\\
d_{2p-1+2pk}\left|2p^2+2p-2+2p(p+2)k,p\right\rangle &=& \left|2p^2+4p-2+2p(p+2)k,p\right\rangle,\\
d_{2p-1+2pk}\left|2p^2+4p-2+2p(p+2)k,p\right\rangle &=& \left|2p^2+6p-2+2p(p+2)k,3p\right\rangle.
\eea
The new states with $\Lambda=3p$ also have $f$- and $d$-descendants, and so on, all of which can be summarized in a diagram, shown in Figure \ref{fig:ContinuousASpecial}.

Starting with the black circle at the bottom left, representing the primary, the next two states on the same row are singular and must be subtracted, so they are colored red.  The first state on the next row has two antecendants, one red and one black.  Thus it has already been added once and subtracted once and has the correct coefficient of zero, so we color it blue to indicate no explicit contribution to the character.  As you continue to move up and to the right, every state has equal numbers of red and black antecedants, and so gets colored blue, indicating no explicit term in the character.  Only the states on the first row contribute, and their contribution is exactly the same as for the generic ($x\ne 1/p$) continuous representation.  So the character ends up the same as before, which is consistent with the analysis of \cite{Gepner:2001px}.  Note that in the third row, each state has two $d$-descendants.  These states also have the property that every state on this row is an $f$-descendant of each state to its left and has every state to its right as an $f$-descendant (we denote them with squares rather than circles or diamonds to indicate this fact).  In the figure there are no $g$-descendants, which would be represented by vertical lines, but this is only because we haven't made the diagram large enough.  Eventually, as we move to the right on any given row, the states will be $g$-descendants of states above them, but again this will not change our conclusion or our expression for the character.

The $Y$ series representations are very similar, with a non-generic point at $x=1/p$, but where the final result is the same character as at the generic point.  The $Z$ series never has any $d$- or $g$-descendants.

\subsection{Alternative expressions}
\label{subsec:NewExpressions}

Now we will give an alternative representation for these continuous representation characters.  The equivalence is conjectural, but we have verified it to very high order in powers of $q$.

Consider two integers $a$ and $b$ satisfying $1\le a\le p+1$, $1\le b\le p-1$.  Then define
\be
\label{eq:ContinuousUCharacter}
\chi[U^{(p)}_{a,b;x}](q)=\frac{q^{x-y_{a,b}}}{\eta(q)}\sum_{k=1}^p\chi^{(p+1)}_{k,a}(q)\chi^{(p)}_{b,k}(q),
\ee
where 
\be
y_{a,b}=\frac{\lp p+2-2a\rp^2}{8p(p+2)}+\frac{\d_{a,1}+\d_{a,p+1}}{p},
\ee
and where we use the equivalences
\be
\chi^{(p+1)}_{k,a}(q)=\chi^{(p+1)}_{p+1-k,p+2-a}(q),\qquad\chi^{(p)}_{b,k}(q)=\chi^{(p)}_{p-b,p+1-k}(q),
\ee
to ensure that the first index is always greater than or equal to the second index on the minimal model characters.

We conjecture that these characters are equivalent to the continuous upper series characters we obtained above using the embedding formalism.  For $a+b$ even they correspond to NS sector characters, while for $a+b$ odd they correspond to the R sector.  The explicit correspondence to the labeling of representations above into the classes of~\cite{Gepner:2001px} is
\begin{align}
X^{(p)}_x=\ & U^{(p)}_{1,1;x},\\
Y^{(p)}_{n;x}=\ & U^{(p)}_{1,2n-1;x},\\
Z^{(p)}_{n,m;x}=\ & U^{(p)}_{2m-n,n;x},\\
V^{(p)}_{n;x}=\ & U^{(p)}_{1,2n;x},\\
W^{(p)}_{n,m;x}=\ & U^{(p)}_{2m-n+1,n;x}.
\end{align}

Note that there is a redundancy in the labeling,
\be
\chi[U^{(p)}_{a,b;x}](q)=\chi[U^{(p)}_{p+2-a,p-b;x}](q).
\ee
Because of this equivalence, there are a total of $(p^2-1)/2$ distinct representations for $p$ odd, and $p^2/2$ for $p$ even (in the latter case, there is one self-dual representation in the R sector, with $a=1+p/2$, $b=p/2$).  For all $p$, half of the representations are NS and half are R.

Since we have
\be
h^{(p+1)}_{k,a}+h^{(p)}_{b,k}=\hlf\lp k-\frac{a+b}{2}\rp^2+\frac{\lp pa-(p+2)b\rp^2-4}{8p(p+2)},
\ee
then for NS the lowest weight state in the representation comes from the $k=\tfrac{a+b}{2}$ term, while for R we have two lowest weight states, from the terms with $k=\tfrac{a+b\pm 1}{2}$.  For NS, the lowest weight state has
\be
a=h^{(p)}_{b,\tfrac{a+b}{2}},\quad h=x+\frac{(a-b)^2p+4+4a-2a^2+2b^2-8\d_{a,1}-8\d_{a,p+1}}{8p},
\ee
and for R they have
\be
a_\pm=h^{(p)}_{b,\tfrac{a+b\pm 1}{2}},\quad h=x+\frac{(a-b)^2p+4+4a-2a^2+2b^2-8\d_{a,1}-8\d_{a,p+1}}{8p}+\frac{1}{8}.
\ee

\subsection{Modular properties}

The characters of the $p$th minimal model, labeled by $1\le m\le n<p$ transform amongst themselves under the S-transformation ($\tau\rr -1/\tau$) of the modular group according to the matrix
\be
S^{(p)}_{n,m;N,M}=\sqrt{\frac{8}{p(p+1)}}\lp -1\rp^{(n+m)(N+M)}\sin(\pi\frac{nN}{p})\sin(\pi\frac{mM}{p+1}).
\ee

We can thus use this result to check the transformations of the $\chi^{(p+1)}\chi^{(p)}$ pieces of $\chi[U^{(p)}_{a,b;x}]$.  We find
\begin{multline}
\label{eq:STransf}
\sum_{k=1}^p\chi^{(p+1)}_{k,a}\chi^{(p)}_{b,k}\longrightarrow\sum_{k=1}^p\sum_{N=1}^p\sum_{M=1}^N\sum_{n=1}^{p-1}\sum_{m=1}^nS^{(p+1)}_{k,a;N,M}S^{(p)}_{b,k;n,m}\chi^{(p+1)}_{N,M}\chi^{(p)}_{n,m}\\
=\frac{1}{4}\sum_{k=1}^p\sum_{N=1}^p\sum_{M=1}^{p+1}\sum_{n=1}^{p-1}\sum_{m=1}^pS^{(p+1)}_{k,a;N,M}S^{(p)}_{b,k;n,m}\chi^{(p+1)}_{N,M}\chi^{(p)}_{n,m}\\
=\frac{2}{(p+1)\sqrt{p(p+2)}}\sum_{k=1}^p\sum_{N=1}^p\sum_{M=1}^{p+1}\sum_{n=1}^{p-1}\sum_{m=1}^p\lp -1\rp^{(N+M)(a+k)+(n+m)(b+k)}\\
\times\sin(\frac{\pi Nk}{p+1})\sin(\frac{\pi Ma}{p+2})\sin(\frac{\pi nb}{p})\sin(\frac{\pi mk}{p+1})\chi^{(p+1)}_{N,M}\chi^{(p)}_{n,m}.
\end{multline}
Next we'll use the result that
\be
\sum_{k=1}^p\lp -1\rp^{(n+m+N+M)k}\sin(\frac{\pi Nk}{p+1})\sin(\frac{\pi mk}{p+1})=\left\{\begin{matrix}\frac{p+1}{2}\d_{N,m}, & N+M+n+m\ \mathrm{even},\\ -\frac{p+1}{2}\d_{N,p+1-m}, & N+M+n+m\ \mathrm{odd}.\end{matrix}\right.
\ee
This converts the sums in (\ref{eq:STransf}) to
\begin{multline}
\frac{1}{\sqrt{p(p+2)}}\sum_{M=1}^{p+1}\sum_{n=1}^{p-1}\sum_{m=1}^p\sin(\frac{\pi Ma}{p+2})\sin(\frac{\pi nb}{p})\\
\times\left\{\frac{1+(-1)^{M+n}}{2}\lp -1\rp^{aM+bn+(a+b)m}\chi^{(p+1)}_{m,M}\chi^{(p)}_{n,m}\right.\\
\left. -\frac{1+(-1)^{M+n+p}}{2}\lp -1\rp^{a(p+1-M)+bn+(a+b)m}\chi^{(p+1)}_{p+1-m,M}\chi^{(p)}_{n,m}\right\}.
\end{multline}

Let's focus separately on the case where $a+b$ is even (NS sector) or odd (R sector).  In the NS sector, this becomes
\begin{multline}
\frac{1}{\sqrt{p(p+2)}}\sum_{A=1}^{p+1}\sum_{B=1}^{p-1}\sum_{k=1}^p\sin(\frac{\pi Aa}{p+2})\sin(\frac{\pi Bb}{p})\\
\times\left\{\frac{1+(-1)^{A+B}}{2}\lp -1\rp^{aA+bB}\chi^{(p+1)}_{k,A}\chi^{(p)}_{B,k}\right.\\
\left. -\frac{1+(-1)^{(p+2-A)+B}}{2}\lp -1\rp^{a(p+1+A)+bB}\chi^{(p+1)}_{k,p+2-A}\chi^{(p)}_{B,k}\right\}\\
=\frac{2}{\sqrt{p(p+2)}}\sum_{A=1}^{p+1}\sum_{B=1}^{p-1}\frac{1+(-1)^{A+B}}{2}\sin(\frac{\pi Aa}{p+2})\sin(\frac{\pi Bb}{p})\sum_{k=1}^p\chi^{(p+1)}_{k,A}\chi^{(p)}_{B,k}.
\end{multline}
In particular, we see that the NS sector closes on itself under S-transformations!  As an example, for $p=3$, we can take a basis $a=b=1$ and $a=b=2$ for the NS states, and then the S-matrix is (keeping in mind an extra factor of two because the sum is over the doubled set of states),
\be
S=\frac{4}{\sqrt{15}}\lp\begin{matrix}\sin(\frac{\pi}{5})\sin(\frac{\pi}{3}) & \sin(\frac{2\pi}{5})\sin(\frac{2\pi}{3}) \\ \sin(\frac{2\pi}{5})\sin(\frac{2\pi}{3}) & \sin(\frac{4\pi}{5})\sin(\frac{4\pi}{3})\end{matrix}\rp=\lp\begin{matrix}\frac{2}{\sqrt{5}}\xi & \hlf\xi^{-1} \\ \hlf\xi^{-1} & -\frac{2}{\sqrt{5}}\xi\end{matrix}\rp,
\ee
where
\be
\xi=\sin(\frac{\pi}{5})=\frac{\sqrt{5-\sqrt{5}}}{2\sqrt{2}}.
\ee
For $p=4$ we can take a basis $\{(a,b)\}=\{(1,1),(1,3),(2,2),(3,1)\}$, and the S-matrix becomes
\be
S=\lp\begin{matrix}\tfrac{1}{2\sqrt{3}} & \tfrac{1}{2\sqrt{3}} & \tfrac{1}{\sqrt{2}} & \tfrac{1}{\sqrt{3}} \\ \tfrac{1}{2\sqrt{3}} & \tfrac{1}{2\sqrt{3}} & -\tfrac{1}{\sqrt{2}} & \tfrac{1}{\sqrt{3}} \\ \tfrac{1}{\sqrt{2}} & -\tfrac{1}{\sqrt{2}} & 0 & 0 \\ \tfrac{1}{\sqrt{3}} & \tfrac{1}{\sqrt{3}} & 0 & -\tfrac{1}{\sqrt{3}}\end{matrix}\rp.
\ee

If we instead start with a R sector state, we get a similar answer,
\begin{multline}
\frac{2}{\sqrt{p(p+2)}}\sum_{A=1}^{p+1}\sum_{B=1}^{p-1}\frac{1+(-1)^{A+B}}{2}\lp -1\rp^A\sin(\frac{\pi Aa}{p+2})\sin(\frac{\pi Bb}{p})\sum_{k=1}^p\lp -1\rp^k\chi^{(p+1)}_{k,A}\chi^{(p)}_{B,k}.
\end{multline}
The result only involves states where $A+B$ is even, i.e.\ only NS sector states, but it is not quite the characters themselves; there is an extra $(-1)^k$ in the sum.  This is consistent with the fact that the R sector partition function is expected, under the S-transformation, to map into the NS sector partition function with an insertion of $(-1)^F$.

\subsection{Lower series characters}

Because it is so difficult to find simple (i.e.\ with nonnegative integer coefficients) combinations of minimal model characters which are closed under S-transformations, we conjecture that nearly the same expressions will apply for the massive characters of the $c<6$ series.  In particular, we conjecture
\be
\label{eq:LSCCBilinearForm}
\chi[\wtU^{(p)}_{a,b;x}](q)=\frac{q^{x-y'_{a,b}}}{\eta(q)}\sum_{k=1}^{p-1}\chi^{(p-1)}_{b,k}(q)\chi^{(p)}_{k,a}(q),
\ee
where
\be
1\le a\le p,\quad 1\le b\le p-2,\qquad y'_{a,b}=\frac{(p-1-2b)^2}{8(p-1)(p+1)},
\ee
and
\be
\chi[\wtU^{(p)}_{a,b;x}](q)=\chi[\wtU^{(p)}_{p+1-a,p-1-b;x}](q).
\ee

\section{Threshold relations}
\label{sec:ThresholdRelations}

For each of the continuous representations, if we approach the lower limit of the $h$ range by sending $x\rr 0$, then some descendant states become null.  Equivalently, we can say that certain states in the representation space are no longer descendants of the primary state; the representation splits into smaller ones.  This decomposition is called a threshold relation.

In terms of characters, it means that the $x\rr 0$ limit of the continuous character is equal to a sum of other characters.  The leading (i.e.\ lowest weight) term in this sum must be a discrete representation character, while subleading terms can a priori be either continuous (but at specific values of $x$) or discrete.  For example, the standard Virasoro algebra with $c>1$ has a continuous family of representations with $h=x>0$ and characters $\chi^{Vir}_x=\eta^{-1}q^{h-\tfrac{c-1}{24}}$, and a single discrete representation with $h=0$ and character $\chi^{Vir}_{vac}=\eta^{-1}q^{-\tfrac{c-1}{24}}(1-q)$.  In that case we have
\be
\lim_{x\rr 0}\chi^{Vir}_x(\tau)=\chi^{Vir}_{vac}(\tau)+\chi^{Vir}_1(\tau).
\ee
For the algebras studied in this paper, we find (or conjecture) that the threshold relations always have the form of a continuous representation decomposing into a pair of discrete representations (unlike the Virasoro case above).

For each of the continuous representations, we know that the first term in the threshold relation must be the uniquely determined discrete representation with the appropriate quantum numbers.  We conjecture that the only additional term in the threshold relation will be the discrete representations corresponding to the first $d$ descendant of the lowest weight state.

For the upper series NS sector, this procedure gives
\begin{align}
\label{eq:ACThresholdRelation}
\lim_{x\rr 0}\chi[X^{(p)}_x]=\ & \lim_{x\rr 0}\chi[U^{(p)}_{1,1;x}]=\chi[A^{(p)}]+\chi[C^{(p)}_{1,2}],\\
\lim_{x\rr 0}\chi[Y^{(p)}_{n;x}]=\ & \lim_{x\rr 0}\chi[U^{(p)}_{1,2n-1;x}]=\chi[D^{(p)}_{2n-1,n}]+\chi[B^{(p)}_{n-1}],\\
\label{eq:CDThresholdRelation}
\lim_{x\rr 0}\chi[Z^{(p)}_{n,m;x}]=\ & \lim_{x\rr 0}\chi[U^{(p)}_{2m-n,n;x}]=\chi[C^{(p)}_{n,m}]+\chi[D^{(p)}_{p-n,p-m}].
\end{align}

The lower series NS sector has the simple relation
\be
\lim_{x\rr 0}\chi[\widetilde{X}^{(p)}_{n,m}]=\lim_{x\rr 0}\chi[\wtU^{(p)}_{m,2n-m;x}]=\chi[\widetilde{A}_{n,m}]+\chi[\widetilde{B}_{n+1,m}].
\ee

In the R sector, for the upper series we have
\begin{align}
\lim_{x\rr 0}\chi[V^{(p)}_{1;x}]=\ & \lim_{x\rr 0}\chi[U^{(p)}_{1,2;x}]=\chi[E^{(p)}]+\chi[F^{(p)}_2],\\
\lim_{x\rr 0}\chi[V^{(p)}_{n;x}]=\ & \lim_{x\rr 0}\chi[U^{(p)}_{1,2n;x}]=\chi[H^{(p)}_{p-2n,p-n}]+\chi[I^{(p)}_n],\\
\lim_{x\rr 0}\chi[W^{(p)}_{n,n;x}]=\ & \lim_{x\rr 0}\chi[U^{(p)}_{n+1,n;x}]=\chi[F^{(p)}_n]+\chi[F^{(p)}_{p-n}],\\
\lim_{x\rr 0}\chi[W^{(p)}_{n,m;x}]=\ & \lim_{x\rr 0}\chi[U^{(p)}_{2m-n+1,n;x}]=\chi[G^{(p)}_{n,m}]+\chi[H^{(p)}_{n,m+1}],
\end{align}
while the lower series satisfies
\begin{align}
\lim_{x\rr 0}\chi[\widetilde{V}^{(p)}_{n,n;x}]=\ & \lim_{x\rr 0}\chi[\wtU^{(p)}_{n,n-1;x}]=\chi[\widetilde{C}^{(p)}_n]+\chi[\widetilde{C}^{(p)}_{p+1-n}],\\
\lim_{x\rr 0}\chi[\widetilde{V}^{(p)}_{n,m;x}]=\ & \lim_{x\rr 0}\chi[\wtU^{(p)}_{m,n-m-1;x}]=\chi[\widetilde{E}^{(p)}_{n,m}]+\chi[\widetilde{D}^{(p)}_{n,m}]
\end{align}

\section{Discrete representation characters}
\label{sec:DiscreteRepCharacters}

\subsection{Upper series $D$ class characters}

Now let's apply this formalism to the discrete representations.  We will only work in detail for the upper series NS sector, and then simply present the results for the other cases.  Let's start with just the $D$ class.  Because of the threshold relations, if we can successfully construct the characters for the $D$ representations, then we can obtain all other characters by combining this information with the continuous representation characters derived above.  Specifically, we would have
\bea
\label{eq:CfromD}
\chi[C^{(p)}_{n,m}] &=& \chi[U^{(p)}_{2m-n,n;0}]-\chi[D^{(p)}_{p-n,p-m}],\\
\label{eq:BfromD}
\chi[B^{(p)}_n] &=& \chi[U^{(p)}_{1,2n+1;0}]-\chi[D^{(p)}_{2n+1,n+1}],\\
\label{eq:AfromD}
\chi[A^{(p)}] &=& \chi[U^{(p)}_{1,1;0}]-\chi[U^{(p)}_{3,1;0}]+\chi[D^{(p)}_{p-1,p-2}].
\eea
Of course, as cross-checks we can try to compute the left-hand sides directly, and we can also try to check that the results are sensible, for instance that they have only positive coefficients in their $q$-expansions.

The $D^{(p)}_{n,m}$ series representations are labeled by a pair of integers,
\be
2\le n\le p-1,\qquad\frac{n}{2}\le m\le n-1.
\ee
Taking $a(D^{(p)}_{n,m})$ and $h(D^{(p)}_{n,m})$ from (\ref{eq:DQuantumNumbers}), we can translate into $\Phi$ and $\La$,
\be
\Phi_0=2\lp p+1\rp n-2pm,\qquad\La_0=4m-2n+p+2.
\ee
It will turn out that for certain values of the combination $2m-n$, which lies in the range
\be
0\le 2m-n\le p-3,
\ee
there will be extra descendants which need to be considered (this will be analogous to the situation with $x=1/p$ in the continuous representations).  So to start, we will assume that $2m-n$ does not take on any of the values $0,1,2,\frac{p+2}{2}$, $\frac{p+6}{2}$.  We will argue that, like in the $x=1/p$ case, the results we obtain are also applicable to these special cases.

First we look for the $f$ descendants of the primary state.  This generates two infinite sequences of descendants,
\be
\left|2p(p+2)j+2n+2pm,\La_0\right\rangle,\qquad\left|2p(p+2)(j+1)-2n-2pm,\La_0\right\rangle,\qquad j\ge 0.
\ee
Let's introduce some shorthand notation, defining
\be
\Phi_{k;a,b,c}=2p(p+2)k+2an+2bm+2c,\qquad \La_j=\left|\La_0+2pj\right|.
\ee
so for instance $\Phi_0=\Phi_{0;p+1,-p,0}$, and the descendants above are
\be
\left|\Phi_{j;1,p,0},\La_0\right\rangle,\qquad\left|\Phi_{j+1;-1,-p,0},\La_0\right\rangle.
\ee
Each of these states in turn has an infinite set of $f$ descendants,
\be
\left|\Phi_{j;p+1,-p,0},\La_0\right\rangle,\qquad\left|\Phi_{j+1;-p-1,p,0},\La_0\right\rangle,\qquad j\ge 0.
\ee
Note that for the first of these, $j=0$ represents the primary state itself.  In the $\Phi\La$-plane, these states all lie on the horizontal line extending rightwards from the primary.

Each of the states so far has a $d$ descendant (and by our genericity restriction, only a single $d$ descendant),
\bea
d_{2pj+2m+1}\left|\Phi_{j;1,p,0},\La_0\right\rangle &=& \left|\Phi_{k;1,p,p},\La_{-1}\right\rangle,\\
d_{2pj-2m+2p-1}\left|\Phi_{j+1;-1,-p,0},\La_0\right\rangle &=& \left|\Phi_{j+1;-1,-p,p},\La_1\right\rangle,\\
d_{2pj+2n-2m-1}\left|\Phi_{j;p+1,-p,0},\La_0\right\rangle &=& \left|\Phi_{j;p+1,-p,p},\La_1\right\rangle,\\
d_{2pj-2n+2m+2p+1}\left|\Phi_{j+1;-p-1,p,0},\La_0\right\rangle &=& \left|\Phi_{j+1;-p-1,p,p},\La_{-1}\right\rangle.
\eea
These new states in turn generate new $f$ descendants.  Some of these are among the $d$ descendants already enumerated, and some are new,
\be
\left|\Phi_{j;1,p,-p},\La_1\right\rangle,\qquad\left|\Phi_{j+1;-p-1,p,-p},\La_1\right\rangle,\qquad j\ge 0,
\ee
\be
\left|\Phi_{j+2;-1,-p,-p},\La_{-1}\right\rangle,\qquad\left|\Phi_{j+1;p+1,-p,-p},\La_{-1}\right\rangle,\qquad j\ge 0.
\ee
And each state also has one $d$ descendant.  For the states that were already obtained as $d$ descendants themselves, their descendants will lie further along the same diagonal line.  For the new states the $d$ descendant diagonals head in the opposite direction, back toward the primary state's horizontal line.  And we could continue in this way, with new $d$ descendants generating new rows and corresponding $f$ descendants ad infinitum.

Note that we have not said anything yet about $g$ descendants.  In fact, we are going to make the assumption, which we will attempt to justify a posteriori, that $g$ descendants do not need to be accounted for when computing these generic $D$ series characters.  This could be, for instance, if the $g$-descendants were not actually linearly independent of the $f$ and $d$ descendant nodes with the same $\Phi$ and $\La$ (it would be interesting to attempt to verify this conjecture by explicit computations).  The other issue with $g$ descendants is that they seem to have significantly more sensitivity to the specific values of $p$, $n$, and $m$, and if their contribution did need to be included it is unlikely that there would be nice expressions for whole families of characters of the type that we will present below.

\begin{figure}
\begin{subfigure}{0.5\textwidth}
\centering
\begin{tikzpicture}[scale=0.35]
   \pgfsetplotmarksize{2.5pt}
   \draw[->] (-1,0) -- (20,0) node[right] {$\Phi$};
   \draw[->] (0,-1) -- (0,10) node[above] {$\Lambda$};
   \draw[dashed] (-0.5,-1) -- (5,10);
   \draw (2.2,2.8) -- (20,2.8);
   \draw (2.2,2.8) -- (5.8,10);
   \draw (3.8,2.8) -- (7.4,10);
   \draw (4.6,5.2) -- (7.2,0) -- (12.2,10);
   \draw (6.2,5.2) -- (8.8,0) -- (13.8,10);
   \draw (9.4,2) -- (10.4,0) -- (15.4,10);
   \draw (11,2) -- (12,0) -- (17,10);
   \draw (11.8,10) -- (16.8,0) -- (20,6.4);
   \draw (13.4,10) -- (18.4,0) -- (20,3.2);
   \draw (15,10) -- (20,0);
   \draw (16.6,10) -- (20,3.2);
   \node at (2.2,2.8) {\pgfuseplotmark{*}};
   \node at (11.8,2.8) {\pgfuseplotmark{*}};
   \node[color=red] at (3.8,2.8) {\pgfuseplotmark{diamond*}};
   \node[color=red] at (13.4,2.8) {\pgfuseplotmark{diamond*}};
   \node[color=blue] at (5.8,2.8) {\pgfuseplotmark{diamond*}};
   \node[color=blue] at (15.4,2.8) {\pgfuseplotmark{diamond*}};
   \node[color=blue] at (7.4,2.8) {\pgfuseplotmark{*}};
   \node[color=blue] at (17,2.8) {\pgfuseplotmark{*}};
   \draw (3.4,5.2) -- (20,5.2);
   \node[color=red] at (3.4,5.2) {\pgfuseplotmark{diamond*}};
   \node[color=red] at (13,5.2) {\pgfuseplotmark{diamond*}};
   \node[color=blue] at (4.6,5.2) {\pgfuseplotmark{*}};
   \node[color=blue] at (14.2,5.2) {\pgfuseplotmark{*}};
   \node at (5,5.2) {\pgfuseplotmark{*}};
   \node at (14.6,5.2) {\pgfuseplotmark{*}};
   \node[color=blue] at (6.2,5.2) {\pgfuseplotmark{diamond*}};
   \node[color=blue] at (15.8,5.2) {\pgfuseplotmark{diamond*}};
   \draw (4.6,7.6) -- (20,7.6);
   \node at (4.6,7.6) {\pgfuseplotmark{*}};
   \node at (14.2,7.6) {\pgfuseplotmark{*}};
   \node[color=red] at (6.2,7.6) {\pgfuseplotmark{diamond*}};
   \node[color=red] at (15.8,7.6) {\pgfuseplotmark{diamond*}};
   \node[color=blue] at (13,7.6) {\pgfuseplotmark{diamond*}};
   \node[color=blue] at (14.6,7.6) {\pgfuseplotmark{*}};
   \draw (5.8,10) -- (20,10);
   \node[color=red] at (5.8,10) {\pgfuseplotmark{diamond*}};
   \node[color=red] at (15.4,10) {\pgfuseplotmark{diamond*}};
   \node at (7.4,10) {\pgfuseplotmark{*}};
   \node at (17,10) {\pgfuseplotmark{*}};
   \node[color=blue] at (11.8,10) {\pgfuseplotmark{*}};
   \node[color=blue] at (13.4,10) {\pgfuseplotmark{diamond*}};
   \draw (7,0.4) -- (20,0.4);
   \node[color=red] at (7,0.4) {\pgfuseplotmark{*}};
   \node[color=red] at (16.6,0.4) {\pgfuseplotmark{*}};
   \node at (8.6,0.4) {\pgfuseplotmark{diamond*}};
   \node at (18.2,0.4) {\pgfuseplotmark{diamond*}};
   \node[color=blue] at (10.6,0.4) {\pgfuseplotmark{diamond*}};
   \node[color=blue] at (12.2,0.4) {\pgfuseplotmark{*}};
   \draw (8.2,2) -- (20,2);
   \node at (8.2,2) {\pgfuseplotmark{diamond*}};
   \node at (17.8,2) {\pgfuseplotmark{diamond*}};
   \node[color=blue] at (9.4,2) {\pgfuseplotmark{*}};
   \node[color=blue] at (19,2) {\pgfuseplotmark{*}};
   \node[color=red] at (9.8,2) {\pgfuseplotmark{*}};
   \node[color=red] at (19.4,2) {\pgfuseplotmark{*}};
   \node[color=blue] at (11,2) {\pgfuseplotmark{diamond*}};
   \draw (9.4,4.4) -- (20,4.4);
   \node[color=red] at (9.4,4.4) {\pgfuseplotmark{*}};
   \node[color=red] at (19,4.4) {\pgfuseplotmark{*}};
   \node at (11,4.4) {\pgfuseplotmark{diamond*}};
   \node[color=blue] at (17.8,4.4) {\pgfuseplotmark{diamond*}};
   \node[color=blue] at (19.4,4.4) {\pgfuseplotmark{*}};
   \draw (10.6,6.8) -- (20,6.8);
   \node at (10.6,6.8) {\pgfuseplotmark{diamond*}};
   \node at (20,6.8) {\pgfuseplotmark{diamond*}};
   \node[color=red] at (12.2,6.8) {\pgfuseplotmark{*}};
   \node[color=blue] at (16.6,6.8) {\pgfuseplotmark{*}};
   \node[color=blue] at (18.2,6.8) {\pgfuseplotmark{diamond*}};
   \draw (11.8,9.2) -- (20,9.2);
   \node[color=red] at (11.8,9.2) {\pgfuseplotmark{*}};
   \node at (13.4,9.2) {\pgfuseplotmark{diamond*}};
   \node[color=blue] at (15.4,9.2) {\pgfuseplotmark{diamond*}};
   \node[color=blue] at (17,9.2) {\pgfuseplotmark{*}};
   \draw[color=gray] (1.2,3.3) -- (2,2.9);
   \node at (1,3.4) {1};
   \draw[color=gray] (2.4,5.7) -- (3.2,5.3);
   \node at (2.2,5.6) {2};
   \draw[color=gray] (3.6,8.1) -- (4.4,7.7);
   \node at (3.4,8) {3};
   \draw[color=gray] (1.2,6.9) -- (4.4,5.3);
   \node at (1,7) {4};
   \draw[color=gray] (3.8,1.8) -- (3.8,2.6);
   \node at (3.8,1.6) {5};
   \draw[color=gray] (5.1,5.3) -- (6.7,6.4);
   \node at (6.9,6.5) {6};
   \draw[color=gray] (5.7,2.7) -- (5.2,2.2);
   \node at (5,2) {7};
   \draw[color=gray] (6.9,0.4) -- (5.4,0.6);
   \node at (5.2,0.6) {8};
\end{tikzpicture}
\caption{}
\label{fig:DiscreteDGeneric}
\end{subfigure}
\begin{subfigure}{0.5\textwidth}
\centering
\begin{tikzpicture}[scale=0.35]
   \pgfsetplotmarksize{2.5pt}
   \draw[->] (-1,0) -- (20,0) node[right] {$\Phi$};
   \draw[->] (0,-1) -- (0,10) node[above] {$\Lambda$};
   \draw[dashed] (-0.5,-1) -- (5,10);
   \draw (1,0.4) -- (20,0.4);
   \draw (1,0.4) -- (5.8,10);
   \draw (2.6,0.4) -- (7.4,10);
   \draw (4.6,5.2) -- (7.2,0) -- (12.2,10);
   \draw (6.2,5.2) -- (8.8,0) -- (13.8,10);
   \draw (9.4,2) -- (10.4,0) -- (15.4,10);
   \draw (11,2) -- (12,0) -- (17,10);
   \draw (11.8,10) -- (16.8,0) -- (20,6.4);
   \draw (13.4,10) -- (18.4,0) -- (20,3.2);
   \draw (15,10) -- (20,0);
   \draw (16.6,10) -- (20,3.2);
   \node at (1,0.4) {\pgfuseplotmark{*}};
   \node at (10.6,0.4) {\pgfuseplotmark{*}};
   \node[color=red] at (2.6,0.4) {\pgfuseplotmark{diamond*}};
   \node[color=red] at (12.2,0.4) {\pgfuseplotmark{diamond*}};
   \node[color=blue] at (7,0.4) {\pgfuseplotmark{diamond*}};
   \node[color=blue] at (16.6,0.4) {\pgfuseplotmark{diamond*}};
   \node[color=blue] at (8.6,0.4) {\pgfuseplotmark{*}};
   \node[color=blue] at (18.2,0.4) {\pgfuseplotmark{*}};
   \draw (2.2,2.8) -- (20,2.8);
   \node[color=red] at (2.2,2.8) {\pgfuseplotmark{diamond*}};
   \node[color=red] at (11.8,2.8) {\pgfuseplotmark{diamond*}};
   \node at (3.8,2.8) {\pgfuseplotmark{*}};
   \node at (13.4,2.8) {\pgfuseplotmark{*}};
   \node[color=blue] at (5.8,2.8) {\pgfuseplotmark{*}};
   \node[color=blue] at (15.4,2.8) {\pgfuseplotmark{*}};
   \node[color=blue] at (7.4,2.8) {\pgfuseplotmark{diamond*}};
   \node[color=blue] at (17,2.8) {\pgfuseplotmark{diamond*}};
   \draw (3.4,5.2) -- (20,5.2);
   \node at (3.4,5.2) {\pgfuseplotmark{*}};
   \node at (13,5.2) {\pgfuseplotmark{*}};
   \node[color=blue] at (4.6,5.2) {\pgfuseplotmark{diamond*}};
   \node[color=blue] at (14.2,5.2) {\pgfuseplotmark{diamond*}};
   \node[color=red] at (5,5.2) {\pgfuseplotmark{diamond*}};
   \node[color=red] at (14.6,5.2) {\pgfuseplotmark{diamond*}};
   \node[color=blue] at (6.2,5.2) {\pgfuseplotmark{*}};
   \node[color=blue] at (15.8,5.2) {\pgfuseplotmark{*}};
   \draw (4.6,7.6) -- (20,7.6);
   \node[color=red] at (4.6,7.6) {\pgfuseplotmark{diamond*}};
   \node[color=red] at (14.2,7.6) {\pgfuseplotmark{diamond*}};
   \node at (6.2,7.6) {\pgfuseplotmark{*}};
   \node at (15.8,7.6) {\pgfuseplotmark{*}};
   \node[color=blue] at (13,7.6) {\pgfuseplotmark{*}};
   \node[color=blue] at (14.6,7.6) {\pgfuseplotmark{diamond*}};
   \draw (5.8,10) -- (20,10);
   \node at (5.8,10) {\pgfuseplotmark{*}};
   \node at (15.4,10) {\pgfuseplotmark{*}};
   \node[color=red] at (7.4,10) {\pgfuseplotmark{diamond*}};
   \node[color=red] at (17,10) {\pgfuseplotmark{diamond*}};
   \node[color=blue] at (11.8,10) {\pgfuseplotmark{diamond*}};
   \node[color=blue] at (13.4,10) {\pgfuseplotmark{*}};
   \draw (8.2,2) -- (20,2);
   \node[color=red] at (8.2,2) {\pgfuseplotmark{*}};
   \node[color=red] at (17.8,2) {\pgfuseplotmark{*}};
   \node[color=blue] at (9.4,2) {\pgfuseplotmark{diamond*}};
   \node[color=blue] at (19,2) {\pgfuseplotmark{diamond*}};
   \node at (9.8,2) {\pgfuseplotmark{diamond*}};
   \node at (19.4,2) {\pgfuseplotmark{diamond*}};
   \node[color=blue] at (11,2) {\pgfuseplotmark{*}};
   \draw (9.4,4.4) -- (20,4.4);
   \node at (9.4,4.4) {\pgfuseplotmark{diamond*}};
   \node at (19,4.4) {\pgfuseplotmark{diamond*}};
   \node[color=red] at (11,4.4) {\pgfuseplotmark{*}};
   \node[color=blue] at (17.8,4.4) {\pgfuseplotmark{*}};
   \node[color=blue] at (19.4,4.4) {\pgfuseplotmark{diamond*}};
   \draw (10.6,6.8) -- (20,6.8);
   \node[color=red] at (10.6,6.8) {\pgfuseplotmark{*}};
   \node[color=red] at (20,6.8) {\pgfuseplotmark{*}};
   \node at (12.2,6.8) {\pgfuseplotmark{diamond*}};
   \node[color=blue] at (16.6,6.8) {\pgfuseplotmark{diamond*}};
   \node[color=blue] at (18.2,6.8) {\pgfuseplotmark{*}};
   \draw (11.8,9.2) -- (20,9.2);
   \node at (11.8,9.2) {\pgfuseplotmark{diamond*}};
   \node[color=red] at (13.4,9.2) {\pgfuseplotmark{*}};
   \node[color=blue] at (15.4,9.2) {\pgfuseplotmark{*}};
   \node[color=blue] at (17,9.2) {\pgfuseplotmark{diamond*}};
\end{tikzpicture}
\caption{}
\label{fig:DiscreteCGeneric}
\end{subfigure}
\caption{The $\Phi\Lambda$ embedding diagrams for generic $D$-series and $C$-series characters (we have stretched the vertical axis by a factor of two to make the diagram more legible).  Specifically, (a) is for $D^{(6)}_{5,4}$ and (b) is for $C^{(6)}_{1,2}$.  The shapes (circles and diamonds) allow us to keep track of direct descendants; all direct descendants of a circle node are diamonds, and vice versa.  The color scheme is to keep track of contributions to the character; black nodes contribute with a plus sign, red nodes contribute with a minus sign, and blue nodes do not contribute to the character.  Note that if we superpose the two diagrams together, then everything above the bottom row cancels, leaving us precisely with a continuous representation diagram as in Figure \ref{fig:ContinuousA}, consistent with the threshold relation (\ref{eq:CDThresholdRelation}).}
\end{figure}
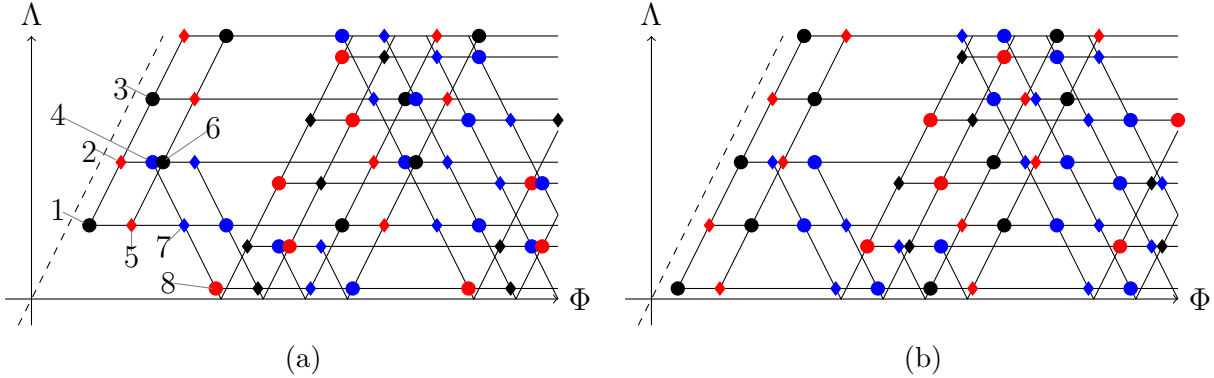

We plot the embedding diagram for this representation in Figure \ref{fig:DiscreteDGeneric}.  It remains to determine how each of these states contribute to the character, i.e.\ to determine the color assignments in the diagram with black nodes representing a plus contribution, red nodes a minus contribution, and blue nodes no contribution (i.e.\ a coefficient of zero).  

For example, the furthest left node, labeled 1, represents the primary state, and is of course colored black.  Correspondingly the leading term in the character would be $P_{NS}(\tau)q^{h-\frac{c}{24}}$, accounting for the primary and all its descendants.  Its $d$ descendant is the diamond up and right from the primary, labeled node 2.  Since this is a null state, it should not be counted by the character, but it was included in the primary's contribution (from expanding the $P_{NS}(\tau)$ factor), so we need to subtract it off with $-P_{NS}(\tau)q^{h'-\frac{c}{24}}$, where $h'$ is the weight of this descendant.  To indicate this, we color the node red.  The $d$ descendant of this state, node 3, is not actually a descendant of the primary, because successive $d$ descendants are nilpotent.  So it has been subtracted from the last contribution and currently appears with a negative sign in the character.  So we need to add it back in, and hence color it black.  This pattern continues up the diagonal, with alternating black circles and red diamonds.  

Consider node 4, to the right of node 2.  As with all the nodes in the diagram, it is a null state and should not appear in the character.  Fortunately, it was already included in the contribution from the primary, node 1, with a plus sign, and in the contribution from node 2, with a minus sign.  Thus it has already been correctly removed and gets colored blue.

The diamond immediately to the right of the primary, node 5, needs to be subtracted off and becomes red.  Its $d$ descendant, node 6, has been added once, from the node 1 contribution, and subtracted twice, as as a descendant of nodes 2 and 5.  So we need to add it back in, and thus color it black.  A similar pattern continues up the diagonal with alternating red and black.

As a couple of final examples, consider nodes 7 and 8.  Node 7 is an $f$ descendant of node 1 and a $d$ descendant of node 4, which in turn was an $f$ descendant of node 2.  Since node 1 is black, node 4 is blue, and node 2 is red, we have a net zero contribution, and node 7 gets colored blue.  Note that node 7 is not a descendant of node 5, since they are both diamonds.  Now for node 8, it is a $d$ descendant of node 7, which is an $f$ descendant of node 1.  We do not count it as being a descendant of nodes 4 or 2 because the route from 4 to 8 involves taking a $d$ descendant twice, which is disallowed (since $d$ descendants correspond to fermionic operators).  Thus we only have to consider the contributions of nodes 7 (blue) and 1 (black), and to cancel these, node 8 must be colored red.

Proceeding in the same manner allows us to fill in all of the colors in Figure \ref{fig:DiscreteDGeneric}.  From here it is straightforward to write down the corresponding character itself.  We can organize the nodes in the diagram by diagonals.  Each diagonal eventually alternates between red and black as it moves to the right, possibly after an initial phase of blue.  Each diagonal contributes to the character as a geometric series, and the net expression is
\begin{align}
\chi[D^{(p)}_{n,m}](q) =\ & P_{NS}(\tau)q^{\frac{pn^2-2(p-1)nm+(p-2)m^2+n-2m+1}{2p}-\frac{1}{4}-\frac{3}{4p}}\sum_{k=0}^\infty q^{\hlf p(p+2)k^2}\left\{\frac{q^{\ls (p+1)n-pm\rs k}}{1+q^{pk+n-m-\hlf}}\right.\non\\
& \left. -\frac{q^{\ls p(p+2)-n-pm\rs k+\hlf\ls p(p+2)-2n-2pm-n^2+2nm\rs}}{1+q^{pk+p-m-\hlf}}-\frac{q^{\ls n+pm+p\rs k+\hlf\ls 1+2m-n^2+2nm\rs}}{1+q^{pk+m+\hlf}}\right.\non\\
& \left. +\frac{q^{\ls p(p+3)-(p+1)n+pm\rs k+\hlf\ls p^2+4p+1-2(p+2)n+2(p+1)m\rs}}{1+q^{pk+p-n+m+\hlf}}\right\}.
\label{eq:DSeriesCharacter}
\end{align}
The factor outside the sum is simply $q^{h-\frac{c}{24}}P_{NS}(\tau)$, representing the primary and all its conformal descendants.  The first term inside the sum, which starts at $1$ when $k=0$, accounts for all diagonals whose alternation starts with a black circle on the same row as the primary (node 1 in Figure \ref{fig:DiscreteDGeneric}).  The next term represents the diagonals that start at a red diamond in the same row (like node 5).  The third term are the diagonals which start at red circles on the row below the primary (node 8 is an example).  And finally the last term in the sum accounts for diagonals starting at a black diamond in that same row (like the node immediately to the right of node 8).

In fact, there is a nice way of rewriting this result so that the sums run over all integers rather than just non-negative ones.  We can combine the two positive terms together and the two negative terms together to obtain
\begin{align}
\chi[D^{(p)}_{n,m}](q) =\ & P_{NS}(\tau)q^{\frac{pn^2-2(p-1)nm+(p-2)m^2+n-2m+1}{2p}-\frac{1}{4}-\frac{3}{4p}}\non\\
& \quad\times\sum_{k\in\Z}q^{\hlf p(p+2)k^2}\left\{\frac{q^{\ls(p+1)n-pm\rs k}}{1+q^{pk+n-m-\hlf}}-\frac{q^{\ls p+n+pm\rs k+\hlf\ls 1+2m-n^2+2nm\rs}}{1+q^{pk+m+\hlf}}\right\}.
\end{align}

We should stress at this point that this character can really only be considered conjectural at this point.  As mentioned above, we have not considered the $g$ descendants at all, and we have also implicitly been assuming that there is a single state corresponding to each node in our diagram, regardless of the path taken to that node, even though it is easy to check that there can be high dimensional spaces of states with the same $\Phi$ and $\La$ quantum numbers.  In the next subsection we will present what evidence we have that these characters are correct.

Before we do that, however, we are going to extend our conjecture even further, and make the guess that, even though we only worked through our derivation for what we called ``generic'' $D$ class characters, that the final result holds even in the non-generic cases.  For example, the unique $D$ character for $p=3$, $D^{(3)}_{2,1}$, would be given, under this assumption by
\be
\chi[D^{(3)}_{2,1}](q)=q^{\hlf}P_{NS}(\tau)\sum_{k\in\Z}^\infty q^{\frac{15}{2}k^2}\left\{\frac{q^{5k}}{1+q^{3k+\hlf}}-\frac{q^{8k+\frac{3}{2}}}{1+q^{3k+\frac{3}{2}}}\right\}.
\ee
This agrees with the results of~\cite{Benjamin:2014kna}, who computed the characters for the $c=12$ theory by similar methods, which lends support to our claim that (\ref{eq:DSeriesCharacter}) is more broadly applicable to all $D$ series characters.

We have also checked several non-generic cases by hand and recovered the same result, for instance the $p=5$ characters computed in Appendix~\ref{app:p5Diagrams}, see in particular Figure~\ref{fig:up5NS}.

\subsection{Other upper series NS characters}

From inserting (\ref{eq:DSeriesCharacter}) and (\ref{eq:ContinuousUCharacter}) into the threshold relations (\ref{eq:CfromD}), (\ref{eq:BfromD}), (\ref{eq:AfromD}) we obtain
\begin{align}
\chi[C^{(p)}_{n,m}] =\ & P_{NS}(q)q^{\frac{pn^2-2(p-1)nm+(p-2)m^2+2m-n+1}{2p}-\frac{1}{4}-\frac{3}{4p}}\non\\
& \quad\times\sum_{k\in\Z}q^{\hlf p(p+2)k^2}\left\{\frac{q^{\ls -(p+1)n+pm\rs k}}{1+q^{pk-n+m-\hlf}}-\frac{q^{\ls n+pm\rs k+\hlf\ls -n^2+2nm\rs}}{1+q^{pk+m-\hlf}}\right\},\\
\chi[B^{(p)}_n] =\ & P_{NS}(q)q^{\frac{(p+2)n^2+2(p+1)n-p}{2p}-\frac{1}{4}-\frac{3}{4p}}\sum_{k\in\Z}q^{\hlf p(p+2)k^2}\left\{\frac{q^{\ls(p+2)n+p+1\rs k}}{1+q^{pk+n-\hlf}}-\frac{q^{\ls(p+2)n+p+1\rs k+1}}{1+q^{pk+n+\frac{3}{2}}}\right\},\\
\chi[A^{(p)}] =\ & P_{NS}(q)q^{-\frac{1}{4}-\frac{3}{4p}}\sum_{k\in\Z}q^{\hlf p(p+2)k^2}\left\{\frac{q^{\ls p+1\rs k-\hlf}}{1+q^{pk-\hlf}}-\frac{q^{\ls p+1\rs k+\hlf}}{1+q^{pk+\frac{3}{2}}}\right\}.
\end{align}

Although we again emphasize that these expressions have not been rigorously derived, and should be treated as conjectural, there are a number of checks that can be done to build our confidence in them.

First of all, we can repeat the direct exercise we did for the $D$ characters in the case of ``generic'' $C$ characters as well.  The corresponding $\Phi\La$-plane embedding diagram is illustrated in Figure \ref{fig:DiscreteCGeneric}.  The result agrees with that obtained from the threshold relation and our previous expression for the continuous spectrum characters.  Similar exercises for the other characters also match the results above, for instance see the $p=5$ diagrams in Figure~\ref{fig:up5NS} in Appendix~\ref{app:p5Diagrams}.  Note that we can no longer ignore the $g$ descendants completely, rather our refined conjecture is that the only time $g$ descendants must be taken into account is when the primary state itself has a $g$ descendant.  An example of this is the $B^{(5)}_1$ representation, whose diagram is shown in Figure~\ref{fig:up5B1}.

Secondly, in the $p=3$ case the expressions are in perfect agreement with the results of~\cite{Benjamin:2014kna}.

Finally, one can check that in the full expressions only non-negative integer coefficients appear, at least up to very high orders in an expansion in powers of $q$.

\subsection{Upper series R sector}

Repeating the procedure for the R sector of the upper series, we can recover characters for the discrete representations.
\begin{align}
\chi[E^{(p)}]=\ & P_R(q)\sum_{k\in\Z}q^{\hlf p(p+2)k^2}\left\{\frac{q^{\hlf(3p+4)k}}{1+q^{pk}}-\frac{q^{\hlf(3p+4)k+1}}{1+q^{pk+2}}\right\},\\
\chi[F^{(p)}_n]=\ & P_R(q)\sum_{k\in\Z}q^{\hlf p(p+2)k^2}\left\{\frac{q^{\hlf (2n+p)k}}{1+q^{pk}}-\frac{q^{((p+1)n+\frac{p}{2})k+\hlf n(n+1)}}{1+q^{pk+n}}\right\},\\
\chi[G^{(p)}_{n,m}]=\ & P_R(q)q^{\frac{2p(m-n)^2-4m(m-n)+2p(m-n)}{4p}}\non\\
& \quad\times\sum_{k\in\Z}q^{\hlf p(p+2)k^2}\left\{\frac{q^{(pm-(p+1)n+\frac{p}{2})k}}{1+q^{pk+m-n}}-\frac{q^{(pm+n+\frac{p}{2})k+\hlf n(2m-n+1)}}{1+q^{pk+m}}\right\},\\
\chi[H^{(p)}_{n,m}]=\ & P_R(q)q^{\frac{2p(m-n)^2+4(m-1)(1-m+n)+2p(m-n)-4p}{4p}}\non\\
& \quad\times\sum_{k\in\Z}q^{\hlf p(p+2)k^2}\left\{\frac{q^{(pm-(p+1)n+\frac{p}{2})k}}{1+q^{pk+m-n-1}}-\frac{q^{(pm+n+\frac{p}{2})k+\hlf n(2m-n+1)}}{1+q^{pk+m-1}}\right\},\\
\chi[I^{(p)}_n]=\ & P_R(q)q^{\frac{p(2n^2+2n-4)+4n^2-4}{4p}}\sum_{k\in\Z}q^{\hlf p(p+2)k^2}\left\{\frac{q^{((p+2)n+\frac{p}{2})k}}{1+q^{pk+n-1}}-\frac{q^{((p+2)n+\frac{p}{2})k+1}}{1+q^{pk+n+1}}\right\}.
\end{align}

A few examples for $p=5$ are shown in Figure~\ref{fig:up5R} in Appendix~\ref{app:p5Diagrams}.  A novel feature that occurs in this case deserves some comment.  For the representations with only a single ground state, there are actually two starting points one can take for the construction of the embedding diagram; corresponding to which direction one moves on the central $\Phi=\La$ diagonal when taking $d_0$ descendants.  The choices produce seeming distinct diagrams (though they agree on the black and red nodes strictly below the diagonal), but the characters are in fact equal, providing a solid check on our procedure.  In Appendix~\ref{app:p5Diagrams}, we include, for the $p=5$ case, these alternative diagrams for the five cases of such representations in Figure~{fig:up5RAlt}.  As in the NS sector case, some of the primaries have $g$ descendants, in particular for $p=5$ one can look at the representations $I^{(5)}_2$ in Figure~\ref{fig:up5I2}, or one of the two possible diagrams for $E^{(5)}$ in Figure~\ref{fig:up5Eb}.

Again, the embedding diagram calculations are in full agreement with the threshold relations, as well as the $p=3$ calculations of~\cite{Benjamin:2014kna}, and again the check that only non-negative coefficients appear in the full characters is another check that our results are sensible.

\subsection{Lower series discrete characters}

Repeating the same procedures once more results in the following characters for the lower series discrete representations.
\begin{align}
\chi[\widetilde{A}^{(p)}_{n,m}]=\ & P_{NS}(q)q^{\frac{p(n-m)^2+(3n-m-1)(n-m+1)-m}{2(p+1)}-\frac{1}{4}+\frac{3}{4(p+1)}}\non\\
& \quad\times\sum_{k\in\Z}q^{\hlf(p-1)(p+1)k^2}\left\{\frac{q^{\ls(p+1)n-pm\rs k}}{1+q^{(p+1)k+n-m+\hlf}}-\frac{q^{\ls(p+1)n-m\rs k+\hlf m(2n-m)}}{1+q^{(p+1)k+n+\hlf}}\right\},\\
\chi[\widetilde{B}^{(p)}_{n,m}]=\ & P_{NS}(q)q^{\frac{p(n-m)^2+(3n-m+1)(n-m-1)+m}{2(p+1)}-\frac{1}{4}+\frac{3}{4(p+1)}}\non\\
& \quad\times\sum_{k\in\Z}q^{\hlf(p-1)(p+1)k^2}\left\{\frac{q^{\ls(p+1)n-pm\rs k}}{1+q^{(p+1)k+n-m-\hlf}}-\frac{q^{\ls(p+1)n-m\rs k+\hlf m(2n-m)}}{1+q^{(p+1)k+n-\hlf}}\right\},\\
\chi[\widetilde{C}^{(p)}_n]=\ & P_R(q)\sum_{k\in\Z}q^{\hlf(p-1)(p+1)k^2}\left\{\frac{q^{\ls n+\frac{p+1}{2}\rs k}}{1+q^{(p+1)k}}-\frac{q^{\ls pn+\frac{p+1}{2}\rs k+\hlf n(n+1)}}{1+q^{(p+1)k+n}}\right\},\\
\chi[\widetilde{D}^{(p)}_{n,m}]=\ & P_R(q)q^{\frac{p(n-m)^2+(n-m)(1-m+3n)+p(n-m)}{2(p+1)}}\non\\
& \quad\times\sum_{k\in\Z}q^{\hlf(p-1)(p+1)k^2}\left\{\frac{q^{\ls(p+1)n-pm+\frac{p+1}{2}\rs k}}{1+q^{(p+1)k+n-m}}-\frac{q^{\ls(p+1)n-m+\frac{p+1}{2}\rs k+\hlf m(2n-m+1)}}{1+q^{(p+1)k+n}}\right\},\\
\chi[\widetilde{E}^{(p)}_{n,m}]=\ & P_R(q)q^{\frac{p(n-m)^2-(n-m)(1+m-3n)-p(n-m)}{2(p+1)}}\non\\
& \quad\times\sum_{k\in\Z}q^{\hlf(p-1)(p+1)k^2}\left\{\frac{q^{\ls(p+1)n-pm-\frac{p+1}{2}\rs k}}{1+q^{(p+1)k+n-m}}-\frac{q^{\ls(p+1)n-m-\frac{p+1}{2}\rs k+\hlf m(2n-m-1)}}{1+q^{(p+1)k+n}}\right\}.
\end{align}

These characters, derived from the embedding diagram formalism, once more satisfy the threshold relations and exhibit only non-negative coefficients in the full $q$ expansion.  One more set of checks is done in the next section, where the $p=3$ results are compared to characters for the supersymmetric free boson and related theories at $c=3/2$.

\section{$c=3/2$ and the supersymmetric free boson}
\label{sec:SpecialCases}

The simplest non-trivial case is $c=3/2$.  Here our system essentially corresponds to a supersymmetric boson $X$ with modes $\al_k$ and its fermionic partner $\psi$ with modes $\psi_r$.  $A(z)$ is simply the stress tensor in the $\psi$ sector, $A(z)=-\hlf:\psi(z)\p\psi(z):$, with $G(z)=\sqrt{2}i\p X(z)\psi(z)$, and $T(z)=A(z)+B(z)$, where $B(z)=-:\p X(z)\p X(x):$ is the stress tensor for $X$ (we take $\al'=1$ here).

In~\cite{Gepner:2001px}, the different representations were identified.  Consider first the NS sector.  The vacuum is of course the $\widetilde{A}^{(3)}_{1,1}$ representation.  To build descendants of the vacuum we can't use modes of $\p X(z)$ or $\psi(z)$ by themselves, but we can use $G(z)$ where every $\al_k$ mode is paired with a $\psi_r$ mode, or we can use $A(z)$ and $B(z)$ which involve pairs of $\psi_r$ or $\al_k$ modes respectively.  For modes built with $\al_k$'s, the free boson includes a family of Virasoro representations $V_n$ with weights $h=n^2$ for $n=0,1,\cdots$, and their corresponding characters are
\be
\chi^{(X)}_{V_n}(z)=\frac{q^{n^2}-q^{(n+1)^2}}{\eta(q)},
\ee
where the states with even $n$ involve even numbers of $\al_{-k}$ oscillators, while the states with odd $n$ involve odd numbers of $\al_{-k}$ oscillators.  This means that the even $n$ states must be paired with $\chi^{(3)}_{1,1}(q)$ from the $\psi$ sector (which includes all NS sector states built from even numbers of $\psi_r$ oscillators), while the odd $n$ states must be paired with $\chi^{(3)}_{2,1}(q)$ (built from odd numbers of $\psi_r$ oscillators).  Schematically we have states $(\sum_{n\ \mathrm{even}}V_n)1+(\sum_{n\ \mathrm{odd}}V_n)\psi$ and the character is
\begin{align}
\chi_1(q)=\ & \sum_{n\ \mathrm{even}}\frac{q^{n^2}-q^{(n+1)^2}}{\eta(q)}\chi^{(3)}_{1,1}(q)+\sum_{n\ \mathrm{odd}}\frac{q^{n^2}-q^{(n+1)^2}}{\eta(q)}\chi^{(3)}_{2,1}(q)\non\\
=\ & \lp\sum_{n=0}^\infty\frac{(-1)^nq^{n^2}}{\eta(q)}\rp\chi^{(3)}_{1,1}(q)+\lp\sum_{n=1}^\infty\frac{(-1)^{n+1}q^{n^2}}{\eta(q)}\rp\chi^{(3)}_{2,1}(q),
\end{align}
Similarly, the operator $\psi(z)$ (or the state $\psi_{-\hlf}|0\rangle$) should correspond to the state $\widetilde{B}^{(3)}_{2,1}$.  Following the same reasoning as above, we find $(\sum_{n\ \mathrm{even}}V_n)\psi+(\sum_{n\ \mathrm{odd}}V_n)1$ with character
\begin{align}
\chi_\psi(q)=\ & \sum_{n\ \mathrm{even}}\frac{q^{n^2}-q^{(n+1)^2}}{\eta(q)}\chi^{(3)}_{2,1}(q)+\sum_{n\ \mathrm{odd}}\frac{q^{n^2}-q^{(n+1)^2}}{\eta(q)}\chi^{(3)}_{1,1}(q)\non\\
=\ & \lp\sum_{n=0}^\infty\frac{(-1)^nq^{n^2}}{\eta(q)}\rp\chi^{(3)}_{2,1}(q)+\lp\sum_{n=1}^\infty\frac{(-1)^{n+1}q^{n^2}}{\eta(q)}\rp\chi^{(3)}_{1,1}(q).
\end{align}
Next, the state $\widetilde{A}^{(3)}_{2,2}$ is meant to correspond to the operator $\s(z)s(z)$, where $\s(z)$ and $s(z)$ are twist fields for the boson and fermion respectively, which both have $h=\frac{1}{16}$.  So this state is in the R sector of $\psi$ and in a sector where the boson has anti-periodic boundary conditions.  We can act on this state with half-integer moded $\al_{-r}$ raising operators, or with integer moded $\psi_{-n}$ operators.  The parity of the two is again correlated (i.e.~we can act with an even number of $\al$'s and an even number of $\psi$'s, or with odd numbers of both), but that just means that we get only one lowest-weight state instead of two.  The character becomes
$$
\chi_{\s s}(q)=\frac{q^{\frac{1}{48}}}{\prod_{n=1}^\infty\lp 1-q^{n-\hlf}\rp}\chi^{(3)}_{2,2}(q).
$$
The NS sector is then rounded out by the bosonic exponentials $:e^{ipX}(z):$, corresponding to the continuous family of representations $\widetilde{X}^{(3)}_{1,1;\frac{p^2}{4}}$.  Since the exponential can ``soak up'' the action of $\al_k$ modes, modes of $G(z)$ can behave as $\psi_r$ modes.  Thus we have characters
\be
\chi_{e^{ipX}}(q)=\frac{q^{\frac{p^2}{4}}}{\eta(q)}\lp\chi^{(3)}_{1,1}(q)+\chi^{(3)}_{2,1}(q)\rp.
\ee

Comparing to our derived expressions, we find agreement
\be
\chi_1(q)=\chi[\widetilde{A}^{(3)}_{1,1}](q),\quad\chi_\psi(q)=\chi[\widetilde{B}^{(3)}_{2,1}](q),\quad\chi_{\s s}(q)=\chi[\widetilde{A}^{(3)}_{2,2}](q),\quad\chi_{e^{ipX}}(q)=\chi[\widetilde{X}^{(3)}_{1,1;\frac{p^2}{4}}](q).
\ee
We have not actually proven that these relations are correct, but we have checked by computer that the series agree up to order $q^{500}$.  Note also that it is easy to check that the threshold relation holds exactly in the form
\be
\lim_{p\rr 0}\chi_{e^{ipX}}(q)=\chi_1(q)+\chi_\psi(q).
\ee

Proceeding similarly in the Ramond sector, Gepner and Noyvert identify three more discrete representations and one continuous representation in terms of free supersymmetric boson constructions.  It will be useful to refer to additional twist fields $\tau(z)$ and $\m(z)$ which appear in the OPEs
\begin{align}
\p X(z)\s(w)\sim\ & \frac{\tau(w)}{(z-w)^\hlf}+\cdots,\\
\psi(z)s(w)\sim\ & \frac{\m(w)}{(z-w)^\hlf}+\cdots,
\end{align}
and where $\tau(z)$ and $\m(z)$ have weights $h=\frac{9}{16}$ and $h=\frac{1}{16}$ respectively. 

The representation $\widetilde{C}^{(3)}_1$ corresponds to the bosonic twist field $\s(z)$.  The descendant states come from acting with $\al_{-r}$ and $\psi_{-s}$ operators, both half-integer moded, with matching parities.  This leads to a character
\begin{align}
\chi_\s(q)=\ & \hlf q^{\frac{1}{48}}\ls\frac{1}{\prod_{n=1}^\infty\lp 1-q^{n-\hlf}\rp}+\frac{1}{\prod_{n=1}^\infty\lp 1+q^{n-\hlf}\rp}\rs\chi^{(3)}_{1,1}(q)\non\\
& \quad +\hlf q^{\frac{1}{48}}\ls\frac{1}{\prod_{n=1}^\infty\lp 1-q^{n-\hlf}\rp}-\frac{1}{\prod_{n=1}^\infty\lp 1+q^{n-\hlf}\rp}\rs\chi^{(3)}_{2,1}(q)\non\\
=\ & \hlf q^{\frac{1}{48}}\ls\frac{\chi^{(3)}_{1,1}(q)+\chi^{(3)}_{2,1}(q)}{\prod_{n=1}^\infty\lp 1-q^{n-\hlf}\rp}+\frac{\chi^{(3)}_{1,1}(q)-\chi^{(3)}_{2,1}(q)}{\prod_{n=1}^\infty\lp 1+q^{n-\hlf}\rp}\rs.
\end{align}
Next up is $\widetilde{C}^{(3)}_2$ corresponding to the fermionic twist field $s(z)$.  In this case we schematically end up with $(\sum_{n\ \mathrm{even}}V_n)s+(\sum_{n\ \mathrm{odd}}V_n)\m$ and the character is
\begin{align}
\chi_s(q)=\ & \sum_{n\ \mathrm{even}}\frac{q^{n^2}-q^{(n+1)^2}}{\eta(q)}\chi^{(3)}_{2,2}(q)+\sum_{n\ \mathrm{odd}}\frac{q^{n^2}-q^{(n+1)^2}}{\eta(q)}\chi^{(3)}_{2,2}(q)\non\\
=\ & \frac{1}{\eta(q)}\chi^{(3)}_{2,2}(q).
\end{align}
The third discrete representation in the R sector is $\widetilde{E}^{(3)}_{2,1}$.  This is a doublet state whose components are $\tau(z)$ and $\s(z)\psi(z)$, both with $h=\frac{9}{16}$.  The corresponding character is similar to $\chi_\s$,
\begin{align}
\chi_{\lp\begin{smallmatrix}\tau \\ \s\psi \end{smallmatrix}\rp}(q)=\ & \hlf q^{\frac{1}{48}}\ls\frac{1}{\prod_{n=1}^\infty\lp 1-q^{n-\hlf}\rp}-\frac{1}{\prod_{n=1}^\infty\lp 1+q^{n-\hlf}\rp}\rs\chi^{(3)}_{1,1}(q)\non\\
& \quad +\hlf q^{\frac{1}{48}}\ls\frac{1}{\prod_{n=1}^\infty\lp 1-q^{n-\hlf}\rp}+\frac{1}{\prod_{n=1}^\infty\lp 1+q^{n-\hlf}\rp}\rs\chi^{(3)}_{2,1}(q)\non\\
=\ & \hlf q^{\frac{1}{48}}\ls\frac{\chi^{(3)}_{1,1}(q)+\chi^{(3)}_{2,1}(q)}{\prod_{n=1}^\infty\lp 1-q^{n-\hlf}\rp}-\frac{\chi^{(3)}_{1,1}(q)-\chi^{(3)}_{2,1}(q)}{\prod_{n=1}^\infty\lp 1+q^{n-\hlf}\rp}\rs.
\end{align}
Finally, the continuous representation $\widetilde{V}^{(3)}_{2,2;\frac{p^2}{4}}$ is also a identified as the doublet with components $:s(z)e^{ipX}(z):$ and $:\m(z)e^{ipX}(z):$, and character
\be
\chi_{\lp\begin{smallmatrix} se^{ipX} \\ \m e^{ipX} \end{smallmatrix}\rp}(q)=2\frac{q^{\frac{p^2}{4}}}{\eta(q)}\chi^{(3)}_{2,2}(q).
\ee
Again the threshold relation works precisely, with
\be
\lim_{p\rr 0}\chi_{\lp\begin{smallmatrix} se^{ipX} \\ \m e^{ipX} \end{smallmatrix}\rp}(q)=2\chi_s(z).
\ee
As in the NS sector, we have confirmed numerically that these characters agree with our expressions up to order $q^{500}$.

\section*{Acknowledgements}

We would like to thank O.~Lunin for useful conversations.
This research was supported by
NSF grant PHY-1820867.

\appendix

\section{Mode algebra}
\label{app:ModeAlgebra}

The chiral algebra of an $\mathcal{SW}(3/2,2)$ algebra is generated by spin-two bosonic currents $T$ and $A$, and fermionic currents $G$ and $M$ of spin three-halves and five-halves respectively.  Performing a standard mode expansion, the algebra is given by
\bea
\ls L_m,L_n\rs &=& \frac{c}{12}m\lp m-1\rp\lp m+1\rp\d_{m+n}+\lp m-n\rp L_{m+n},\\
\ls L_m,G_r\rs &=& \lp\frac{m}{2}-r\rp G_{m+r},\\
\{G_r,G_s\} &=& \frac{c}{3}\lp r-\hlf\rp\lp r+\hlf\rp\d_{r+s}+2L_{r+s},\\
\ls L_m,A_n\rs &=& \frac{c(15-c)}{36(12+c)}m\lp m-1\rp\lp m+1\rp\d_{m+n}+\lp m-n\rp A_{m+n},\\
\ls A_m,G_r\rs &=& -\frac{15-c}{2(12+c)}\lp r+\hlf\rp G_{m+r}-M_{m+r},\\
\label{eq:AACommutator}
\ls A_m,A_n\rs &=& \frac{c(15-c)}{36(12+c)}m\lp m-1\rp\lp m+1\rp\d_{m+n}+\lp m-n\rp A_{m+n},\\
\ls L_m,M_r\rs &=& \frac{15-c}{4(12+c)} m\lp m+1\rp G_{m+r}+\lp\frac{3m}{2}-r\rp M_{m+r},\\
\{G_r,M_s\} &=& \frac{c(15-c)}{18(12+c)}\lp r+\hlf\rp\lp r-\hlf\rp\lp r-\frac{3}{2}\rp\d_{r+s}\non\\
&& \quad -\frac{15-c}{12+c}\lp r+\hlf\rp L_{r+s}+\lp 3r-s\rp A_{r+s},\\
\ls A_m,M_r\rs &=& \frac{15-c}{4(12+c)^2}\ls 3\lp 3+c\rp\lp m+1\rp\lp r+\frac{3}{2}\rp\right.\non\\
&& \qquad\left. +\lp 15-c\rp\lp m+r+\frac{5}{2}\rp\lp m+r+\frac{3}{2}\rp\rs G_{m+r}\\
&& \quad +\ls\lp m+1\rp+\frac{15-c}{2(12+c)}\lp m+r+\frac{5}{2}\rp\rs M_{m+r}-\frac{18}{12+c}:AG:_{m+r},\non\\
\{M_r,M_s\} &=& -\frac{c(15-c)}{36(12+c)}\lp r+\frac{3}{2}\rp\lp r+\hlf\rp\lp r-\hlf\rp\lp r-\frac{3}{2}\rp\d_{r+s}\non\\
&& \quad +\frac{(15-c)(3+c)}{2(12+c)^2}\ls 3\lp r+\frac{3}{2}\rp\lp s+\frac{3}{2}\rp\right.\non\\
&& \qquad\left. \vphantom{\lp\frac{3}{2}\rp}-\lp r+s+3\rp\lp r+s+2\rp\rs L_{r+s}+\ls 2\lp r+\frac{3}{2}\rp\lp s+\frac{3}{2}\rp\right.\non\\
&& \qquad\left. -\frac{3+c}{2(12+c)}\lp r+s+3\rp\lp r+s+2\rp\rs A_{r+s}\non\\
&& \quad -\frac{36}{12+c}:TA:_{r+s}+\frac{18}{12+c}:GM:_{r+s},
\eea
where $m$ and $n$ are integers, and $r$ and $s$ are either integers, in the Ramond sector, or integers plus a half, in the Neveu-Schwarz sector.

We have used the definition
\be
:PQ:_n=\sum_{m\le -h_P}P_mQ_{n-m}+(-1)^{PQ}\sum_{m\ge -h_P+1}Q_{n-m}P_m,
\ee
which works for all cases except $:GM:_{r+s}$ in the Ramond sector, where we have (translating results from Appendix C of~\cite{Gepner:2001px} into our conventions),
\be
:GM:^R_n=\sum_{m\le -1}G_mM_{n-m}-\sum_{m\ge 0}M_{n-m}G_m+\frac{5c\lp 15-c\rp}{384\lp 12+c\rp}\d_{n,0}+\frac{3+n}{2}A_n-\frac{15-c}{8\lp 12+c\rp}L_n.
\ee

The Hermitian conjugates for $L$, $G$, and $A$ are standard,
\be
(L_n)^\dagger=L_{-n},\quad(G_r)^\dagger=G_{-r},\quad(A_n)^\dagger=A_{-n},
\ee
while
\be
(M_r)^\dagger=-M_{-r}-\frac{15-c}{2(12+c)}G_{-r}.
\ee

\section{Embedding diagrams for $c=48/5$}
\label{app:p5Diagrams}

\begin{figure}
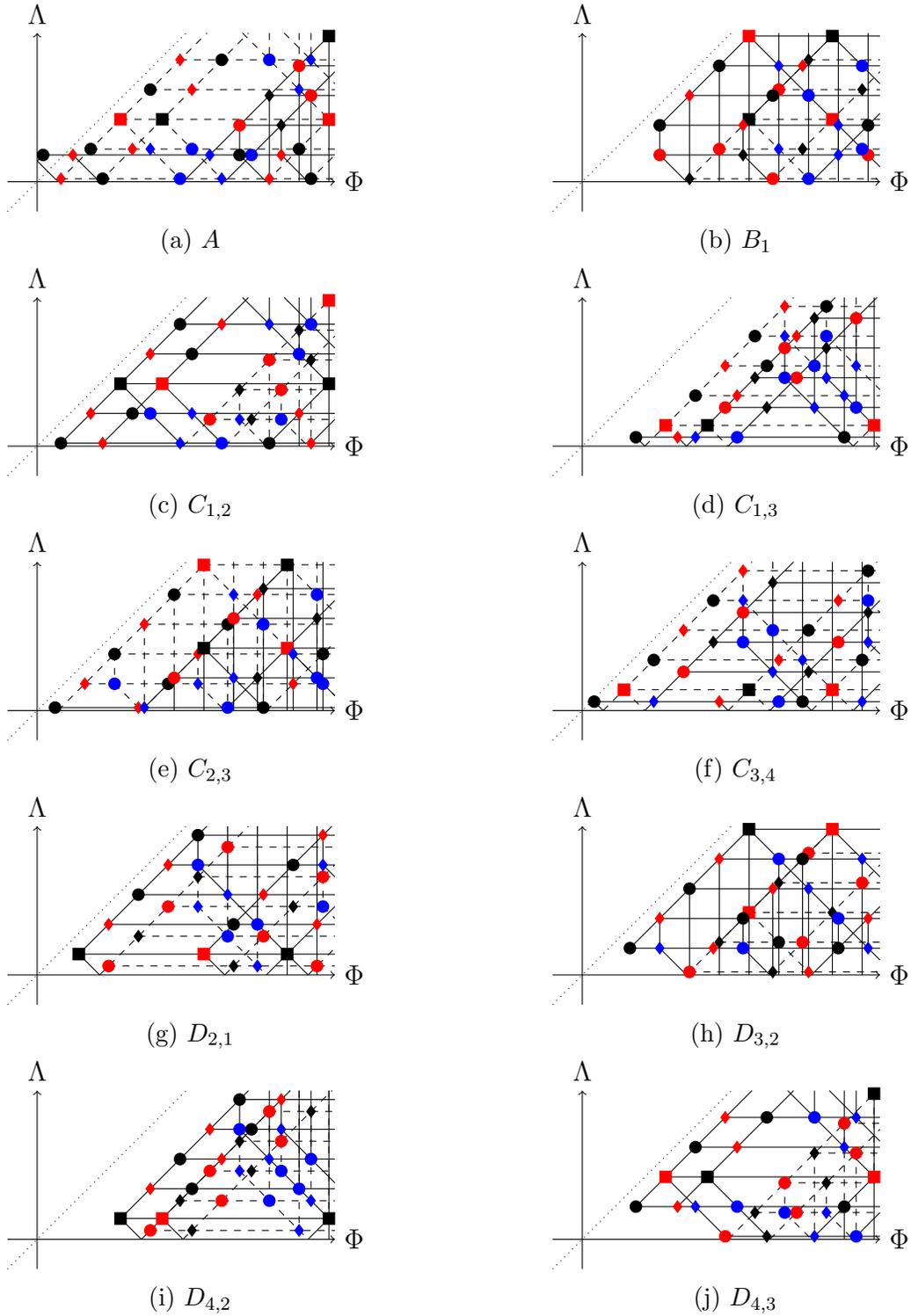

\begin{subfigure}{0.5\textwidth}
\centering

\caption{$A$}
\label{fig:up5A}
\end{subfigure}
\begin{subfigure}{0.5\textwidth}
\centering
%
\caption{$B_1$}
\label{fig:up5B1}
\end{subfigure}
\begin{subfigure}{0.5\textwidth}
\centering
%
\caption{$C_{1,2}$}
\label{fig:up5C12}
\end{subfigure}
\begin{subfigure}{0.5\textwidth}
\centering
%
\caption{$C_{1,3}$}
\label{fig:up5C13}
\end{subfigure}
\begin{subfigure}{0.5\textwidth}
\centering
%
\caption{$C_{2,3}$}
\label{fig:up5C23}
\end{subfigure}
\begin{subfigure}{0.5\textwidth}
\centering
%
\caption{$D_{2,1}$}
\label{fig:up5D21}
\end{subfigure}
\begin{subfigure}{0.5\textwidth}
\centering
%
\caption{$D_{3,2}$}
\label{fig:up5D32}
\end{subfigure}
\begin{subfigure}{0.5\textwidth}
\centering
%
\caption{$D_{4,2}$}
\label{fig:up5D42}
\end{subfigure}
\begin{subfigure}{0.5\textwidth}
\centering
%
\caption{$D_{4,3}$}
\label{fig:up5D43}
\end{subfigure}
\caption{Embedding diagrams for the NS sector discrete characters of the $c=48/5$ (upper series, $p=5$) theory.}
\label{fig:up5NS}
\end{figure}

In this appendix we present embedding diagrams for all of the discrete representations of the upper series with $p=5$.  This case has a central charge of $c=48/5$ and has twenty discrete representations, ten each in the NS and R sectors.  The ten representations in the NS sector are shown in Figure~\ref{fig:up5NS}.  In these diagrams we indicate all $f$ descendants by horizontal lines, all $g$ descendants by vertical lines, and all $d$ descendants by diagonal lines.  We use dashed lines for lines that have ``reflected'' off the $\La=0$ axis (in order to make the images slightly more visually coherent).  The coloring of nodes indicates their contribution to the character: black nodes must be added in, red nodes need to be subtracted out, and blue nodes don't need to be adjusted.

\begin{figure}
\begin{subfigure}{0.5\textwidth}
\centering
%
\caption{$E$}
\label{fig:up5E}
\end{subfigure}
\begin{subfigure}{0.5\textwidth}
\centering
%
\caption{$F_1$}
\label{fig:up5F1}
\end{subfigure}
\begin{subfigure}{0.5\textwidth}
\centering
%
\caption{$F_2$}
\label{fig:up5F2}
\end{subfigure}
\begin{subfigure}{0.5\textwidth}
\centering
%
\caption{$F_3$}
\label{fig:up5F3}
\end{subfigure}
\begin{subfigure}{0.5\textwidth}
\centering
%
\caption{$G_{1,2}$}
\label{fig:up5G12}
\end{subfigure}
\begin{subfigure}{0.5\textwidth}
\centering
%
\caption{$G_{2,3}$}
\label{fig:up5G23}
\end{subfigure}
\begin{subfigure}{0.5\textwidth}
\centering
%
\caption{$H_{1,3}$}
\label{fig:up5H13}
\end{subfigure}
\begin{subfigure}{0.5\textwidth}
\centering
%
\caption{$H_{2,4}$}
\label{fig:up5H24}
\end{subfigure}
\begin{subfigure}{0.5\textwidth}
\centering
%
\caption{$I_2$}
\label{fig:up5I2}
\end{subfigure}
\caption{Embedding diagrams for the R sector discrete characters of the $c=48/5$ (upper series, $p=5$) theory.}
\label{fig:up5R}
\end{figure}

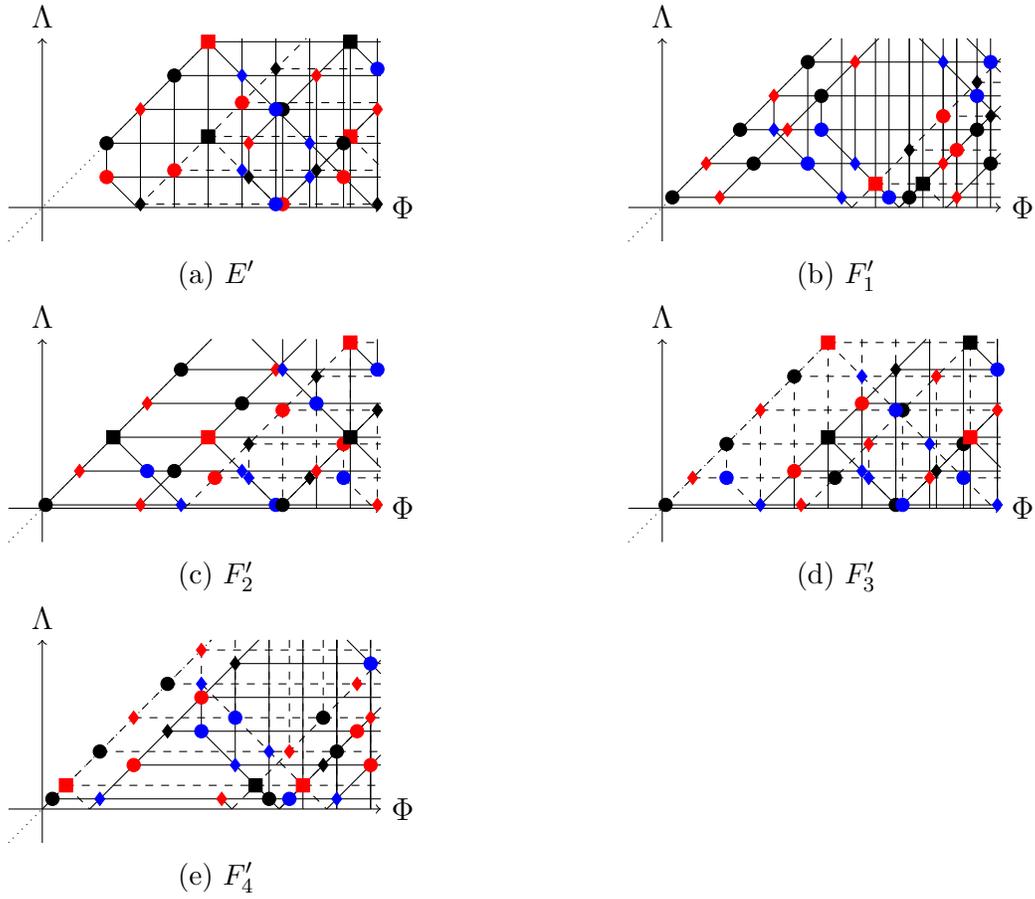
\begin{figure}
\begin{subfigure}{0.5\textwidth}
\centering
\begin{tikzpicture}[scale=0.45]
   \pgfsetplotmarksize{2.5pt}
   \draw[->] (-1,0) -- (10,0) node[right] {$\Phi$};
   \draw[->] (0,-1) -- (0,5) node[above] {$\Lambda$};
   \draw[dotted] (-1,-1) -- (5,5);
   \draw (1.9,1.9) -- (10,1.9);
   \draw (1.9,1.9) -- (5,5);
   \draw (1.9,0.9) -- (2.8,0);
   \draw[dashed] (2.8,0) -- (7.8,5);
   \draw (6.1,1.9) -- (9.2,5);
   \draw (6.1,0.9) -- (7,0);
   \draw[dashed] (7,0) -- (10,3);
   \draw[dashed] (4.9,2.1) -- (7,0);
   \draw (7,0) -- (10,3);
   \draw (4.9,4.9) -- (9.8,0);
   \draw[dashed] (9.8,0) -- (10,0.2);
   \draw[dashed] (9.1,2.1) -- (10,1.2);
   \draw (9.1,4.9) -- (10,4);
   \draw (2.9,2.9) -- (10,2.9);
   \draw (3.9,3.9) -- (10,3.9);
   \draw (4.9,4.9) -- (10,4.9);
   \draw (1.9,0.9) -- (10,0.9);
   \draw[dashed] (2.9,0.1) -- (10,0.1);
   \draw[dashed] (3.9,1.1) -- (10,1.1);
   \draw[dashed] (4.9,2.1) -- (10,2.1);
   \draw[dashed] (5.9,3.1) -- (10,3.1);
   \draw[dashed] (6.9,4.1) -- (10,4.1);
   \draw (1.9,1.9) -- (1.9,0.9);
   \draw (6.1,1.9) -- (6.1,0.9);
   \draw (7.9,1.9) -- (7.9,0.9);
   \draw (8.9,1.9) -- (8.9,0.9);
   \draw (2.9,2.9) -- (2.9,0);
   \draw (6.9,2.9) -- (6.9,0);
   \draw (7.1,2.9) -- (7.1,0);
   \draw (9.9,2.9) -- (9.9,0);
   \draw (3.9,3.9) -- (3.9,0);
   \draw (5.9,3.9) -- (5.9,0);
   \draw (8.1,3.9) -- (8.1,0);
   \draw (4.9,4.9) -- (4.9,0);
   \draw (9.1,4.9) -- (9.1,0);
   \draw (5.9,5) -- (5.9,0);
   \draw (6.9,5) -- (6.9,0);
   \draw (7.9,5) -- (7.9,0);
   \draw (8.9,5) -- (8.9,0);
   \draw (9.9,5) -- (9.9,0);
   \node at (1.9,1.9) {\pgfuseplotmark{*}};
   \node[color=red] at (2.9,2.9) {\pgfuseplotmark{diamond*}};
   \node at (3.9,3.9) {\pgfuseplotmark{*}};
   \node[color=red] at (4.9,4.9) {\pgfuseplotmark{square*}};

   \node[color=red] at (1.9,0.9) {\pgfuseplotmark{*}};
   \node at (2.9,0.1) {\pgfuseplotmark{diamond*}};
   \node[color=red] at (3.9,1.1) {\pgfuseplotmark{*}};
   \node at (4.9,2.1) {\pgfuseplotmark{square*}};
   \node[color=red] at (5.9,3.1) {\pgfuseplotmark{*}};
   \node at (6.9,4.1) {\pgfuseplotmark{diamond*}};

   \node at (6.1,0.9) {\pgfuseplotmark{diamond*}};
   \node[color=red] at (7.1,0.1) {\pgfuseplotmark{*}};
   \node at (8.1,1.1) {\pgfuseplotmark{diamond*}};
   \node[color=red] at (9.1,2.1) {\pgfuseplotmark{square*}};

   \node[color=red] at (6.1,1.9) {\pgfuseplotmark{diamond*}};
   \node at (7.1,2.9) {\pgfuseplotmark{*}};
   \node[color=red] at (8.1,3.9) {\pgfuseplotmark{diamond*}};
   \node at (9.1,4.9) {\pgfuseplotmark{square*}};

   \node[color=blue] at (5.9,3.9) {\pgfuseplotmark{diamond*}};
   \node[color=blue] at (6.9,2.9) {\pgfuseplotmark{*}};
   \node[color=blue] at (7.9,1.9) {\pgfuseplotmark{diamond*}};
   \node[color=red] at (8.9,0.9) {\pgfuseplotmark{*}};
   \node at (9.9,0.1) {\pgfuseplotmark{diamond*}};

   \node[color=blue] at (5.9,1.1) {\pgfuseplotmark{diamond*}};
   \node[color=blue] at (6.9,0.1) {\pgfuseplotmark{*}};
   \node[color=blue] at (7.9,0.9) {\pgfuseplotmark{diamond*}};
   \node at (8.9,1.9) {\pgfuseplotmark{*}};
   \node[color=red] at (9.9,2.9) {\pgfuseplotmark{diamond*}};

   \node[color=blue] at (9.9,4.1) {\pgfuseplotmark{*}};
\end{tikzpicture}
\caption{$E'$}
\label{fig:up5Eb}
\end{subfigure}
\begin{subfigure}{0.5\textwidth}
\centering
\begin{tikzpicture}[scale=0.45]
   \pgfsetplotmarksize{2.5pt}
   \draw[->] (-1,0) -- (10,0) node[right] {$\Phi$};
   \draw[->] (0,-1) -- (0,5) node[above] {$\Lambda$};
   \draw[dotted] (-1,-1) -- (5,5);
   \draw (0.3,0.3) -- (10,0.3);
   \draw (0.3,0.3) -- (5,5);
   \draw (1.7,0.3) -- (6.4,5);
   \draw (3.3,2.3) -- (5.6,0);
   \draw[dashed] (5.6,0) -- (10,4.4);
   \draw (4.7,2.3) -- (7,0);
   \draw[dashed] (7,0) -- (10,3);
   \draw[dashed] (6.3,0.7) -- (7,0);
   \draw (7,0) -- (10,3);
   \draw[dashed] (7.7,0.7) -- (8.4,0);
   \draw (8.4,0) -- (10,1.6);
   \draw (7.6,5) -- (10,2.6);
   \draw (9,5) -- (10,4);
   \draw (1.3,1.3) -- (10,1.3);
   \draw (2.3,2.3) -- (10,2.3);
   \draw (3.3,3.3) -- (10,3.3);
   \draw (4.3,4.3) -- (10,4.3);
   \draw[dashed] (6.3,0.7) -- (10,0.7);
   \draw[dashed] (7.3,1.7) -- (10,1.7);
   \draw[dashed] (8.3,2.7) -- (10,2.7);
   \draw[dashed] (9.3,3.7) -- (10,3.7);
   \draw (3.3,3.3) -- (3.3,2.3);
   \draw (4.7,3.3) -- (4.7,2.3);
   \draw (9.3,3.3) -- (9.3,2.3);
   \draw (4.3,4.3) -- (4.3,1.3);
   \draw (5.7,4.3) -- (5.7,1.3);
   \draw (8.3,4.3) -- (8.3,1.3);
   \draw (9.7,4.3) -- (9.7,1.3);
   \draw (5.3,5) -- (5.3,0.3);
   \draw (6.7,5) -- (6.7,0.3);
   \draw (7.3,5) -- (7.3,0.3);
   \draw (8.7,5) -- (8.7,0.3);
   \draw (6.3,5) -- (6.3,0);
   \draw (7.7,5) -- (7.7,0);
   \draw (7.3,5) -- (7.3,0);
   \draw (8.7,5) -- (8.7,0);
   \draw (8.3,5) -- (8.3,0);
   \draw (9.7,5) -- (9.7,0);
   \draw (9.3,5) -- (9.3,0);
   \node at (0.3,0.3) {\pgfuseplotmark{*}};
   \node[color=red] at (1.3,1.3) {\pgfuseplotmark{diamond*}};
   \node at (2.3,2.3) {\pgfuseplotmark{*}};
   \node[color=red] at (3.3,3.3) {\pgfuseplotmark{diamond*}};
   \node at (4.3,4.3) {\pgfuseplotmark{*}};

   \node[color=red] at (1.7,0.3) {\pgfuseplotmark{diamond*}};
   \node at (2.7,1.3) {\pgfuseplotmark{*}};
   \node[color=red] at (3.7,2.3) {\pgfuseplotmark{diamond*}};
   \node at (4.7,3.3) {\pgfuseplotmark{*}};
   \node[color=red] at (5.7,4.3) {\pgfuseplotmark{diamond*}};

   \node[color=blue] at (3.3,2.3) {\pgfuseplotmark{diamond*}};
   \node[color=blue] at (4.3,1.3) {\pgfuseplotmark{*}};
   \node[color=blue] at (5.3,0.3) {\pgfuseplotmark{diamond*}};
   \node[color=red] at (6.3,0.7) {\pgfuseplotmark{square*}};
   \node at (7.3,1.7) {\pgfuseplotmark{diamond*}};
   \node[color=red] at (8.3,2.7) {\pgfuseplotmark{*}};
   \node at (9.3,3.7) {\pgfuseplotmark{diamond*}};

   \node[color=blue] at (4.7,2.3) {\pgfuseplotmark{*}};
   \node[color=blue] at (5.7,1.3) {\pgfuseplotmark{diamond*}};
   \node[color=blue] at (6.7,0.3) {\pgfuseplotmark{*}};
   \node at (7.7,0.7) {\pgfuseplotmark{square*}};
   \node[color=red] at (8.7,1.7) {\pgfuseplotmark{*}};
   \node at (9.7,2.7) {\pgfuseplotmark{diamond*}};

   \node at (7.3,0.3) {\pgfuseplotmark{*}};
   \node[color=red] at (8.3,1.3) {\pgfuseplotmark{diamond*}};
   \node at (9.3,2.3) {\pgfuseplotmark{*}};

   \node[color=red] at (8.7,0.3) {\pgfuseplotmark{diamond*}};
   \node at (9.7,1.3) {\pgfuseplotmark{*}};

   \node[color=blue] at (8.3,4.3) {\pgfuseplotmark{diamond*}};
   \node[color=blue] at (9.3,3.3) {\pgfuseplotmark{*}};

   \node[color=blue] at (9.7,4.3) {\pgfuseplotmark{*}};
\end{tikzpicture}
\caption{$F_1'$}
\label{fig:up5F1b}
\end{subfigure}
\begin{subfigure}{0.5\textwidth}
\centering
\begin{tikzpicture}[scale=0.45]
   \pgfsetplotmarksize{2.5pt}
   \draw[->] (-1,0) -- (10,0) node[right] {$\Phi$};
   \draw[->] (0,-1) -- (0,5) node[above] {$\Lambda$};
   \draw[dotted] (-1,-1) -- (5,5);
   \draw (0.1,0.1) -- (10,0.1);
   \draw (0.1,0.1) -- (5,5);
   \draw (2.9,0.1) -- (7.8,5);
   \draw (2.1,2.1) -- (4.2,0);
   \draw[dashed] (4.2,0) -- (9.2,5);
   \draw (4.9,2.1) -- (7,0);
   \draw[dashed] (7,0) -- (10,3);
   \draw[dashed] (6.1,0.9) -- (7,0);
   \draw (7,0) -- (10,3);
   \draw [dashed] (8.9,0.9) -- (9.8,0);
   \draw (9.8,0) -- (10,0.2);
   \draw (6.2,5) -- (10,1.2);
   \draw (9,5) -- (10,4);
   \draw (1.1,1.1) -- (10,1.1);
   \draw (2.1,2.1) -- (10,2.1);
   \draw (3.1,3.1) -- (10,3.1);
   \draw (4.1,4.1) -- (10,4.1);
   \draw[dashed] (5.1,0.9) -- (10,0.9);
   \draw[dashed] (6.1,1.9) -- (10,1.9);
   \draw[dashed] (7.1,2.9) -- (10,2.9);
   \draw[dashed] (8.1,3.9) -- (10,3.9);
   \draw[dashed] (9.1,4.9) -- (10,4.9);
   \draw (7.1,5) -- (7.1,4.1);
   \draw (9.9,5) -- (9.9,4.1);
   \draw (8.1,5) -- (8.1,3.1);
   \draw (9.1,5) -- (9.1,2.1);
   \draw[dashed] (6.1,1.9) -- (6.1,0.9);
   \draw[dashed] (8.9,1.9) -- (8.9,0.9);
   \draw[dashed] (7.1,2.9) -- (7.1,0.1);
   \draw (7.1,0.1) -- (7.1,0);
   \draw[dashed] (9.9,2.9) -- (9.9,0.1);
   \draw (9.9,0.1) -- (9.9,0);
   \draw[dashed] (8.1,3.9) -- (8.1,1.1);
   \draw (8.1,1.1) -- (8.1,0);
   \draw[dashed] (9.1,4.9) -- (9.1,2.1);
   \draw (9.1,2.1) -- (9.1,0);
   \node at (0.1,0.1) {\pgfuseplotmark{*}};
   \node[color=red] at (1.1,1.1) {\pgfuseplotmark{diamond*}};
   \node at (2.1,2.1) {\pgfuseplotmark{square*}};
   \node[color=red] at (3.1,3.1) {\pgfuseplotmark{diamond*}};
   \node at (4.1,4.1) {\pgfuseplotmark{*}};

   \node[color=red] at (2.9,0.1) {\pgfuseplotmark{diamond*}};
   \node at (3.9,1.1) {\pgfuseplotmark{*}};
   \node[color=red] at (4.9,2.1) {\pgfuseplotmark{square*}};
   \node at (5.9,3.1) {\pgfuseplotmark{*}};
   \node[color=red] at (6.9,4.1) {\pgfuseplotmark{diamond*}};

   \node[color=blue] at (3.1,1.1) {\pgfuseplotmark{*}};
   \node[color=blue] at (4.1,0.1) {\pgfuseplotmark{diamond*}};
   \node[color=red] at (5.1,0.9) {\pgfuseplotmark{*}};
   \node at (6.1,1.9) {\pgfuseplotmark{diamond*}};
   \node[color=red] at (7.1,2.9) {\pgfuseplotmark{*}};
   \node at (8.1,3.9) {\pgfuseplotmark{diamond*}};
   \node[color=red] at (9.1,4.9) {\pgfuseplotmark{square*}};

   \node[color=blue] at (5.9,1.1) {\pgfuseplotmark{diamond*}};
   \node[color=blue] at (6.9,0.1) {\pgfuseplotmark{*}};
   \node at (7.9,0.9) {\pgfuseplotmark{diamond*}};
   \node[color=red] at (8.9,1.9) {\pgfuseplotmark{*}};
   \node at (9.9,2.9) {\pgfuseplotmark{diamond*}};

   \node[color=blue] at (6.1,0.9) {\pgfuseplotmark{diamond*}};
   \node at (7.1,0.1) {\pgfuseplotmark{*}};
   \node[color=red] at (8.1,1.1) {\pgfuseplotmark{diamond*}};
   \node at (9.1,2.1) {\pgfuseplotmark{square*}};

   \node[color=blue] at (8.9,0.9) {\pgfuseplotmark{*}};
   \node[color=red] at (9.9,0.1) {\pgfuseplotmark{diamond*}};

   \node[color=blue] at (7.1,4.1) {\pgfuseplotmark{diamond*}};
   \node[color=blue] at (8.1,3.1) {\pgfuseplotmark{*}};

   \node[color=blue] at (9.9,4.1) {\pgfuseplotmark{*}};
\end{tikzpicture}
\caption{$F_2'$}
\label{fig:up5F2b}
\end{subfigure}
\begin{subfigure}{0.5\textwidth}
\centering
\begin{tikzpicture}[scale=0.45]
   \pgfsetplotmarksize{2.5pt}
   \draw[->] (-1,0) -- (10,0) node[right] {$\Phi$};
   \draw[->] (0,-1) -- (0,5) node[above] {$\Lambda$};
   \draw[dotted] (-1,-1) -- (5,5);
   \draw (0.1,0.1) -- (10,0.1);
   \draw (0.1,0.1) -- (0,0);
   \draw[dashed] (0,0) -- (5,5);
   \draw[dashed] (1.9,0.9) -- (2.8,0);
   \draw (2.8,0) -- (7.8,5);
   \draw (4.1,0.1) -- (4.2,0);
   \draw[dashed] (4.2,0) -- (9.2,5);
   \draw (4.9,2.1) -- (7,0);
   \draw[dashed] (7,0) -- (10,3);
   \draw[dashed] (6.1,0.9) -- (7,0);
   \draw (7,0) -- (10,3);
   \draw[dashed] (4.9,4.9) -- (9.8,0);
   \draw (9.8,0) -- (10,0.2);
   \draw (9.1,2.1) -- (10,1.2);
   \draw (9,5) -- (10,4);
   \draw (3.9,1.1) -- (10,1.1);
   \draw (4.9,2.1) -- (10,2.1);
   \draw (5.9,3.1) -- (10,3.1);
   \draw (6.9,4.1) -- (10,4.1);
   \draw[dashed] (0.9,0.9) -- (10,0.9);
   \draw[dashed] (1.9,1.9) -- (10,1.9);
   \draw[dashed] (2.9,2.9) -- (10,2.9);
   \draw[dashed] (3.9,3.9) -- (10,3.9);
   \draw[dashed] (4.9,4.9) -- (10,4.9);
   \draw (9.9,5) -- (9.9,4.1);
   \draw[dashed] (1.9,1.9) -- (1.9,0.9);
   \draw[dashed] (6.1,1.9) -- (6.1,0.9);
   \draw[dashed] (7.9,1.9) -- (7.9,0.9);
   \draw[dashed] (8.9,1.9) -- (8.9,0.9);
   \draw[dashed] (2.9,2.9) -- (2.9,0.1);
   \draw (2.9,0.1) -- (2.9,0);
   \draw[dashed] (6.9,2.9) -- (6.9,0.1);
   \draw (6.9,0.1) -- (6.9,0);
   \draw[dashed] (7.1,2.9) -- (7.1,0.1);
   \draw (7.1,0.1) -- (7.1,0);
   \draw[dashed] (9.9,2.9) -- (9.9,0.1);
   \draw (9.9,0.1) -- (9.9,0);
   \draw[dashed] (3.9,3.9) -- (3.9,1.1);
   \draw (3.9,1.1) -- (3.9,0);
   \draw[dashed] (5.9,3.9) -- (5.9,1.1);
   \draw (5.9,1.1) -- (5.9,0);
   \draw[dashed] (8.1,3.9) -- (8.1,1.1);
   \draw (8.1,1.1) -- (8.1,0);
   \draw[dashed] (4.9,4.9) -- (4.9,2.1);
   \draw (4.9,2.1) -- (4.9,0);
   \draw[dashed] (9.1,4.9) -- (9.1,2.1);
   \draw (9.1,2.1) -- (9.1,0);
   \draw[dashed] (5.9,5) -- (5.9,3.1);
   \draw (5.9,3.1) -- (5.9,0);
   \draw[dashed] (6.9,5) -- (6.9,4.1);
   \draw (6.9,4.1) -- (6.9,0);
   \draw[dashed] (9.9,5) -- (9.9,4.1);
   \draw (9.9,4.1) -- (9.9,0);
   \draw (7.9,5) -- (7.9,0);
   \draw (8.9,5) -- (8.9,0);
   \draw (9.9,5) -- (9.9,0);
   \node at (0.1,0.1) {\pgfuseplotmark{*}};
   \node[color=red] at (0.9,0.9) {\pgfuseplotmark{diamond*}};
   \node at (1.9,1.9) {\pgfuseplotmark{*}};
   \node[color=red] at (2.9,2.9) {\pgfuseplotmark{diamond*}};
   \node at (3.9,3.9) {\pgfuseplotmark{*}};
   \node[color=red] at (4.9,4.9) {\pgfuseplotmark{square*}};

   \node[color=blue] at (1.9,0.9) {\pgfuseplotmark{*}};
   \node[color=blue] at (2.9,0.1) {\pgfuseplotmark{diamond*}};
   \node[color=red] at (3.9,1.1) {\pgfuseplotmark{*}};
   \node at (4.9,2.1) {\pgfuseplotmark{square*}};
   \node[color=red] at (5.9,3.1) {\pgfuseplotmark{*}};
   \node at (6.9,4.1) {\pgfuseplotmark{diamond*}};

   \node[color=red] at (4.1,0.1) {\pgfuseplotmark{diamond*}};
   \node at (5.1,0.9) {\pgfuseplotmark{*}};
   \node[color=red] at (6.1,1.9) {\pgfuseplotmark{diamond*}};
   \node at (7.1,2.9) {\pgfuseplotmark{*}};
   \node[color=red] at (8.1,3.9) {\pgfuseplotmark{diamond*}};
   \node at (9.1,4.9) {\pgfuseplotmark{square*}};

   \node[color=blue] at (5.9,1.1) {\pgfuseplotmark{diamond*}};
   \node at (6.9,0.1) {\pgfuseplotmark{*}};
   \node[color=red] at (7.9,0.9) {\pgfuseplotmark{diamond*}};
   \node at (8.9,1.9) {\pgfuseplotmark{*}};
   \node[color=red] at (9.9,2.9) {\pgfuseplotmark{diamond*}};

   \node[color=blue] at (6.1,0.9) {\pgfuseplotmark{diamond*}};
   \node[color=blue] at (7.1,0.1) {\pgfuseplotmark{*}};
   \node at (8.1,1.1) {\pgfuseplotmark{diamond*}};
   \node[color=red] at (9.1,2.1) {\pgfuseplotmark{square*}};

   \node[color=blue] at (5.9,3.9) {\pgfuseplotmark{diamond*}};
   \node[color=blue] at (6.9,2.9) {\pgfuseplotmark{*}};
   \node[color=blue] at (7.9,1.9) {\pgfuseplotmark{diamond*}};
   \node[color=blue] at (8.9,0.9) {\pgfuseplotmark{*}};
   \node[color=blue] at (9.9,0.1) {\pgfuseplotmark{diamond*}};

   \node[color=blue] at (9.9,4.1) {\pgfuseplotmark{*}};
\end{tikzpicture}
\caption{$F_3'$}
\label{fig:up5F3b}
\end{subfigure}
\begin{subfigure}{0.5\textwidth}
\centering
\begin{tikzpicture}[scale=0.45]
   \pgfsetplotmarksize{2.5pt}
   \draw[->] (-1,0) -- (10,0) node[right] {$\Phi$};
   \draw[->] (0,-1) -- (0,5) node[above] {$\Lambda$};
   \draw[dotted] (-1,-1) -- (5,5);
   \draw (0.3,0.3) -- (10,0.3);
   \draw (0.3,0.3) -- (0,0);
   \draw[dashed] (0.3,0.3) -- (5,5);
   \draw[dashed] (0.7,0.7) -- (1.4,0);
   \draw (1.4,0) -- (6.4,5);
   \draw (5.3,0.3) -- (5.6,0);
   \draw[dashed] (5.6,0) -- (10,4.4);
   \draw (4.7,2.3) -- (7,0);
   \draw[dashed] (7,0) -- (10,3);
   \draw[dashed] (6.3,0.7) -- (7,0);
   \draw (7,0) -- (10,3);
   \draw[dashed] (4.7,3.7) -- (8.4,0);
   \draw (8.4,0) -- (10,1.6);
   \draw (9,5) -- (10,4);
   \draw (2.7,1.3) -- (10,1.3);
   \draw (3.7,2.3) -- (10,2.3);
   \draw (4.7,3.3) -- (10,3.3);
   \draw (5.7,4.3) -- (10,4.3);
   \draw[dashed] (0.7,0.7) -- (10,0.7);
   \draw[dashed] (1.7,1.7) -- (10,1.7);
   \draw[dashed] (2.7,2.7) -- (10,2.7);
   \draw[dashed] (3.7,3.7) -- (10,3.7);
   \draw[dashed] (4.7,4.7) -- (10,4.7);
   \draw (4.7,3.3) -- (4.7,2.3);
   \draw (5.7,4.3) -- (5.7,1.3);
   \draw (9.7,4.3) -- (9.7,1.3);
   \draw (6.7,5) -- (6.7,0.3);
   \draw (8.7,5) -- (8.7,0.3);
   \draw (7.7,5) -- (7.7,0);
   \draw (8.7,5) -- (8.7,0);
   \draw (9.7,5) -- (9.7,0);
   \draw[dashed] (4.7,4.7) -- (4.7,3.7);
   \draw[dashed] (5.7,5) -- (5.7,2.7);
   \draw[dashed] (8.3,5) -- (8.3,2.7);
   \draw[dashed] (6.7,5) -- (6.7,1.7);
   \draw[dashed] (7.3,5) -- (7.3,1.7);
   \draw[dashed] (8.7,5) -- (8.7,1.7);
   \draw[dashed] (7.7,5) -- (7.7,0.7);
   \draw[dashed] (8.7,5) -- (8.7,0.3);
   \draw (8.7,0.3) -- (8.7,0);
   \draw[dashed] (9.7,5) -- (9.7,1.3);
   \draw (9.7,1.3) -- (9.7,0);
   \node at (0.3,0.3) {\pgfuseplotmark{*}};
   \node[color=red] at (0.7,0.7) {\pgfuseplotmark{square*}};
   \node at (1.7,1.7) {\pgfuseplotmark{*}};
   \node[color=red] at (2.7,2.7) {\pgfuseplotmark{diamond*}};
   \node at (3.7,3.7) {\pgfuseplotmark{*}};
   \node[color=red] at (4.7,4.7) {\pgfuseplotmark{diamond*}};

   \node[color=blue] at (1.7,0.3) {\pgfuseplotmark{diamond*}};
   \node[color=red] at (2.7,1.3) {\pgfuseplotmark{*}};
   \node at (3.7,2.3) {\pgfuseplotmark{diamond*}};
   \node[color=red] at (4.7,3.3) {\pgfuseplotmark{*}};
   \node at (5.7,4.3) {\pgfuseplotmark{diamond*}};

   \node[color=red] at (5.3,0.3) {\pgfuseplotmark{diamond*}};
   \node at (6.3,0.7) {\pgfuseplotmark{square*}};
   \node[color=red] at (7.3,1.7) {\pgfuseplotmark{diamond*}};
   \node at (8.3,2.7) {\pgfuseplotmark{*}};
   \node[color=red] at (9.3,3.7) {\pgfuseplotmark{diamond*}};

   \node[color=blue] at (4.7,2.3) {\pgfuseplotmark{*}};
   \node[color=blue] at (5.7,1.3) {\pgfuseplotmark{diamond*}};
   \node at (6.7,0.3) {\pgfuseplotmark{*}};
   \node[color=red] at (7.7,0.7) {\pgfuseplotmark{square*}};
   \node at (8.7,1.7) {\pgfuseplotmark{*}};
   \node[color=red] at (9.7,2.7) {\pgfuseplotmark{diamond*}};

   \node[color=blue] at (7.3,0.3) {\pgfuseplotmark{*}};
   \node at (8.3,1.3) {\pgfuseplotmark{diamond*}};
   \node[color=red] at (9.3,2.3) {\pgfuseplotmark{*}};

   \node[color=blue] at (4.7,3.7) {\pgfuseplotmark{diamond*}};
   \node[color=blue] at (5.7,2.7) {\pgfuseplotmark{*}};
   \node[color=blue] at (6.7,1.7) {\pgfuseplotmark{diamond*}};
   \node[color=blue] at (8.7,0.3) {\pgfuseplotmark{diamond*}};
   \node[color=red] at (9.7,1.3) {\pgfuseplotmark{*}};

   \node[color=blue] at (9.7,4.3) {\pgfuseplotmark{*}};
\end{tikzpicture}
\caption{$F_4'$}
\label{fig:up5F4b}
\end{subfigure}
\caption{Equivalent alternative embedding diagrams for the R ground state characters of the $c=48/5$ (upper series, $p=5$) theory.}
\label{fig:up5RAlt}
\end{figure}

Similarly, Figure~\ref{fig:up5R} show the R sector representations.  Five of the ten are Ramond sector ground states (with $h=2/5$ in this case).  For these five, one can give an alternate presentation of the embedding diagram, shown in Figure~\ref{fig:up5RAlt}, but these can be checked to lead to the same expressions for the characters.

\newpage


\providecommand{\href}[2]{#2}

\end{document}